\documentclass[a4paper,11pt,round,twoside]{article}
\usepackage[utf8]{inputenc} 
\usepackage[english]{babel}
\usepackage{hyphenat}
\usepackage{url}
\usepackage{parskip}
\usepackage{enumerate,mdwlist}
\usepackage[usenames,dvipsnames,svgnames,table]{xcolor}
\usepackage{mathtools} 
\usepackage{cases}
\usepackage{empheq}
\usepackage{nameref}
\usepackage{amsmath,amsthm}
\usepackage{amsfonts}
\usepackage{amssymb}
\usepackage{makeidx}
\usepackage{graphicx}
\graphicspath{ {Figures/} }
\usepackage{array}
\usepackage{hyperref}
\usepackage[capposition=top]{floatrow}
\usepackage{subcaption}
\usepackage{xfrac}
\hypersetup{citecolor = Black, colorlinks=true, urlcolor = Black, linkcolor = Black}
\usepackage[left=3.2cm,right=3.2cm,top=3.2cm,bottom=3.2cm]{geometry}
\usepackage{booktabs}
\usepackage{ragged2e}
\usepackage{pgf,tikz}
\usepackage[flushleft]{threeparttable} 
\usetikzlibrary{decorations, positioning, decorations.pathreplacing, calc,arrows.meta}  
\usepackage{mathrsfs}
\usetikzlibrary{arrows}
\usetikzlibrary[patterns]
\usepackage [autostyle, english = american]{csquotes} 
\MakeOuterQuote{"}

\definecolor{lightblueprint}{HTML}{4C9ED9}

\makeatletter
\newcommand{\customlabel}[2]{%
	\protected@write \@auxout {}{\string \newlabel {#1}{{#2}{}}}}
\makeatother

\usepackage[bottom]{footmisc}

\author{
Saara H{\"a}m{\"a}l{\"a}inen\footnote{University of Helsinki and Helsinki GSE, email: \href{mailto:saara.hamalainen@helsinki.fi}{saara.hamalainen@helsinki.fi}.} \, \textcircled{r}  \,  
Eduardo Ferraz\footnote{University of Lille, email: \href{mailto:eduardoferrazmail@gmail.com}{eduardoferrazmail@gmail.com}.} \, \textcircled{r}  \,
Sebasti{\'a}n Castillo Ramos\footnote{University of Helsinki, Helsinki GSE, the Finnish Centre of Excellence in Tax Systems Research (FIT), and WAPLAC. P.O. Box 17, FI-00014 University of Helsinki, FINLAND. Email: \href{mailto:sebastian.castillo@helsinki.fi}{sebastian.castillo@helsinki.fi}}
}
\title{\Large \textbf{Optimal Hybrid  Tax Enforcement}\footnote{We thank Jukka Pirttilä and the participants at the IIPF Annual Congress for helpful comments and suggestions. Sebastián Castillo Ramos acknowledges the support of the Finnish Centre of Excellence in Tax Systems Research funded by the Academy of Finland (project 346251) and the Finnish Cultural Foundation (grant No. 00250253). Saara H{\"a}m{\"a}l{\"a}inen acknowledges the support of the Finnish Research Council (project 360871), the Finnish Cultural Foundation (grant No. 00190358), and Jenny and Antti Wihuri Foundation (grant Nos. 00220099 and 00230118). Any shortcomings are our own.} 
}

\usepackage{setspace}
\usepackage[sort&compress]{natbib}
\usepackage{breqn}
\usepackage{blindtext}
\usepackage{sectsty}

\newtheoremstyle{oma}
 {6pt}   
 {6pt}   
  {\itshape}  
  {12pt}       
  {\bfseries} 
  {}         
  {5pt plus 1pt minus 1pt} 
  {}          

\theoremstyle{oma}

\newtheorem{remark}{Remark}

\newtheorem{lemma}{Lemma}
\newtheorem{proposition}{Proposition}
\newtheorem{corollary}{Corollary}

\newtheorem{example}{Example}

\usepackage{titlesec}

\titleformat*{\section}{\LARGE\bfseries}
\titleformat*{\subsection}{\Large\bfseries}
\titleformat*{\subsubsection}{\large\bfseries}
\titleformat*{\paragraph}{\small\bfseries}
\titleformat*{\subparagraph}{\Small\bfseries}

\tikzset{
pattern size/.store in=\mcSize, 
pattern size = 5pt,
pattern thickness/.store in=\mcThickness, 
pattern thickness = 0.3pt,
pattern radius/.store in=\mcRadius, 
pattern radius = 1pt}
\makeatletter
\pgfutil@ifundefined{pgf@pattern@name@_98mgs5duz}{
\pgfdeclarepatternformonly[\mcThickness,\mcSize]{_98mgs5duz}
{\pgfqpoint{-\mcThickness}{-\mcThickness}}
{\pgfpoint{\mcSize}{\mcSize}}
{\pgfpoint{\mcSize}{\mcSize}}
{
\pgfsetcolor{\tikz@pattern@color}
\pgfsetlinewidth{\mcThickness}
\pgfpathmoveto{\pgfpointorigin}
\pgfpathlineto{\pgfpoint{0}{\mcSize}}
\pgfusepath{stroke}
}}
\makeatother

\tikzset{
pattern size/.store in=\mcSize, 
pattern size = 5pt,
pattern thickness/.store in=\mcThickness, 
pattern thickness = 0.3pt,
pattern radius/.store in=\mcRadius, 
pattern radius = 1pt}
\makeatletter
\pgfutil@ifundefined{pgf@pattern@name@_zl8curl6m}{
\pgfdeclarepatternformonly[\mcThickness,\mcSize]{_zl8curl6m}
{\pgfqpoint{0pt}{0pt}}
{\pgfpoint{\mcSize+\mcThickness}{\mcSize+\mcThickness}}
{\pgfpoint{\mcSize}{\mcSize}}
{
\pgfsetcolor{\tikz@pattern@color}
\pgfsetlinewidth{\mcThickness}
\pgfpathmoveto{\pgfqpoint{0pt}{0pt}}
\pgfpathlineto{\pgfpoint{\mcSize+\mcThickness}{\mcSize+\mcThickness}}
\pgfusepath{stroke}
}}
\makeatother

\tikzset{
pattern size/.store in=\mcSize, 
pattern size = 5pt,
pattern thickness/.store in=\mcThickness, 
pattern thickness = 0.3pt,
pattern radius/.store in=\mcRadius, 
pattern radius = 1pt}
\makeatletter
\pgfutil@ifundefined{pgf@pattern@name@_k3xkycoi0}{
\pgfdeclarepatternformonly[\mcThickness,\mcSize]{_k3xkycoi0}
{\pgfqpoint{-\mcThickness}{-\mcThickness}}
{\pgfpoint{\mcSize}{\mcSize}}
{\pgfpoint{\mcSize}{\mcSize}}
{
\pgfsetcolor{\tikz@pattern@color}
\pgfsetlinewidth{\mcThickness}
\pgfpathmoveto{\pgfpointorigin}
\pgfpathlineto{\pgfpoint{0}{\mcSize}}
\pgfusepath{stroke}
}}
\makeatother

\tikzset{
pattern size/.store in=\mcSize, 
pattern size = 5pt,
pattern thickness/.store in=\mcThickness, 
pattern thickness = 0.3pt,
pattern radius/.store in=\mcRadius, 
pattern radius = 1pt}
\makeatletter
\pgfutil@ifundefined{pgf@pattern@name@cjmdm0tx}{
\pgfdeclarepatternformonly[\mcThickness,\mcSize]{cjmdm0tx}
{\pgfqpoint{0pt}{0pt}}
{\pgfpoint{\mcSize}{\mcSize}}
{\pgfpoint{\mcSize}{\mcSize}}
{
\pgfsetcolor{\tikz@pattern@color}
\pgfsetlinewidth{\mcThickness}
\pgfpathmoveto{\pgfqpoint{0pt}{\mcSize}}
\pgfpathlineto{\pgfpoint{\mcSize+\mcThickness}{-\mcThickness}}
\pgfpathmoveto{\pgfqpoint{0pt}{0pt}}
\pgfpathlineto{\pgfpoint{\mcSize+\mcThickness}{\mcSize+\mcThickness}}
\pgfusepath{stroke}
}}

\begin{document}

\maketitle

\begin{abstract}
\noindent

Modern tax design and enforcement rely on classification data reported by taxpayers. 
Data improve the targeting of tax policy and audits but 
also open avenues for evasion.
How can tax authorities use such data while accounting for their potential strategic distortion?
We demonstrate that restoring deterrence generates an endogenous global verification frontier: 
every report below the frontier is verified through either a profit audit or data verification.
Optimal hybrid tax enforcement requires selecting which data classes are
subject to profit audits up to the global frontier and which are
profit-audited only up to lower class-specific cutoffs, with data
verification taking over beyond these cutoffs.
This costly data verification requirement limits the benefit from differentiated tax enforcement.
Optimal implementation is characterized by profit bunching but involves
no data bunching.

\vspace{0.5cm}

\textbf{Keywords:} \emph{Profit audit, Data verification, Hybrid enforcement policy, Verification frontier, Enforcement index, Bunching}.
\textbf{JEL-codes:} D82, D83, H26, H83. 
\end{abstract}

\newpage

\section{Introduction}

Modern tax systems make extensive use of self-declared data to differentiate tax obligations and enforcement.
Corporate tax returns, for example, contain detailed expense information and balance-sheet items that firms must report under the appropriate categories to claim deductions and exemptions.
Such data can improve the targeting of tax policy and audits, thus mitigating the regressivity bias of tax enforcement \citep{scotchmer1987audit,macho2002auditing,Kuchumova2017}. 
Yet, data-driven tax policies also create natural incentives to distort reported information.\footnote{The Internal Revenue Service (IRS) currently leverages AI to uncover misclassified inputs, such as personal versus business flights and employees versus contractors \citep{gao2026irsai}.}

A growing empirical literature documents widespread manipulation of classifications.
Firms respond to size-dependent enforcement thresholds by adjusting actual or reported revenue or organizational form, generating sharp distortions and bunching patterns \citep{almunia2018under,almunia2022strategic,best2015production,elschner2013special}.
Similarly, a differentiated tax regime across production technologies—such as deductions or exemptions for specific activities—induces misclassification, relabeling of activities, and business splitting \citep{fisman2004tax,chen2021notching,aghion2023tax,kuchumova2025beyond}.
While such findings are consistent with evidence that reporting behavior responds endogenously to the design of tax policy and enforcement \citep{slemrod2007cheating,kleven2011unwilling,fack2016effect,sharma2025optimal}, tax theory largely abstracts from strategic \textit{data} reporting.

In this paper, we study how tax enforcement should optimally account for self-reported data that are both informative and manipulable.
The tax authority has two verification instruments: costly profit audits that verify income and cheaper data audits that detect misclassification and may trigger subsequent profit verification.
The central tradeoff is that, 
while data allow the authority to differentiate enforcement across firms, 
self-classification allows firms to manipulate both enforcement and the tax base.
Preserving truthful data reporting thus places significant limits on differentiation.

Our framework provides a systematic approach to studying the structure of optimal enforcement policies  under strategic data reporting.
We find that restoring deterrence introduces a global \emph{verification frontier}. When enforcement differs across classes, firms assigned to more strongly enforced classes have an incentive to jointly misreport profits and classification data to access more favorable tax treatment. Eliminating these joint deviations requires that every report below the frontier is verified, 
either directly through a profit audit or through the verification of the classification data. Firms reporting the same profit may face different forms of enforcement depending on their data class: a profit report may trigger a profit audit in one class and data verification in another. 

A practical implication of this frontier structure is that optimal enforcement is inherently nonseparable across data classes and cannot be optimized class by class:  differentiated tax treatment generates an endogenous verification requirement that links profit audits and data verification and calls for integration between enforcement functions. For example, the IRS organizes enforcement through specialized units, including Tax Exempt and Government Entities, Large Business and International, and Small Business/Self-Employed, while coordinating across divisions through data-driven workload selection and cross-divisional case classification.
Our results highlight the importance of such collaboration and
efficient allocation of the enforcement budget across departments.\footnote{
This coordination is instrumental when enforcement units have very different resources. Our results show that audit resources cannot be freely shifted between units: reducing profit-audit coverage in one class may call for significant data verification in that class.}

More specifically, our characterization reveals the costs and benefits of differentiated enforcement across data classes.
Building on \citet{sanchez1993hierarchical}, we derive a local \emph{enforcement index} that captures the marginal gains from extending profit audits within each class.
We demonstrate how this relatively overlooked index can be leveraged to assess the efficiency of class-by-class profit-audit allocation under strategic data reporting.\footnote{By contrast, fiscal returns or the marginal value of public funds (MVPF; \citealp{hendrensprungkeyser2020,boning2025welfare}) 
alone are insufficient for determining optimal allocation of audits, which must also account for firms' information rents.
}
The index can be readily estimated from administrative data.\footnote{Empirical implementation requires the tax schedule and the conditional profit distribution within each data class (Appendix~\ref{app:mvpf}).}

The form of optimal enforcement depends on the shape of the enforcement
index. When the index is decreasing in profits (e.g., under a linear tax schedule with a decreasing inverse hazard rate),
enforcement operates on the
\textit{intensive margin}. 
As the audit cutoff rises, 
each additional audit deters underreporting among fewer high-profit firms, 
thus generating diminishing returns and favoring class-specific cutoffs.
We demonstrate that optimal hybrid tax enforcement combines profit audits and data verification, with the mix of instruments varying across classes.
In some classes, 
profit audits stop at class-specific cutoffs, 
where data verification takes over and continues up to the global verification frontier.
In other classes, 
the authority relies exclusively on profit audits, 
thus avoiding the additional costs of data verification on profit reports below the frontier.
These alternative outcomes show how the costs of preserving truthful data reporting can outweigh the benefits of differentiated profit-audit cutoffs, despite their promise for improved audit targeting.\footnote{In our quantitative illustration, this tradeoff results in partial pooling of profit-audit cutoffs across data classes (Appendix~\ref{app:quantitative}).}

When the enforcement index increases in profits (e.g., under a sufficiently
progressive tax schedule), by contrast, enforcement operates on the
\textit{extensive margin}.
As the audit cutoff rises,
the gains from deterring underreporting increase with profits.
Whenever a class is audited,
profit audits thus extend all the way up to the global frontier.
Conditional on the global verification frontier, the enforcement
decision therefore shifts from choosing class-specific audit cutoffs
to deciding which classes are audited for profits up to the frontier and which
are wholly exempt from profit audits. 
Importantly, exemption from profit audits does not imply the absence of
enforcement altogether: these classes still undergo systematic data verification to
prevent firms in profit-audited classes from obtaining weaker enforcement
through misclassification.
The resulting division between profit-audited and data-verified classes echoes the IRS's division-based approach to enforcement and highlights how optimal enforcement may involve fundamentally different verification regimes across taxpayer classes.

Our characterization also provides a diagnostic for whether data verification
is sufficient. Optimal implementation generates bunching in reported profits
at profit-audit cutoffs, but no bunching in classification data: firms respond
to the limits of profit enforcement, whereas verification removes the gain
from strategically changing data classes. Empirical findings of bunching
around size-dependent thresholds and strategic relabeling of activities and
expenses \citep{almunia2018under,chen2021notching} may therefore indicate that
data-contingent policy is not successfully implemented.

Finally, we provide a measure to quantify the value of self-reported data for enforcement.
More informative data improve targeting, but strategic reporting requires
costly verification that diverts resources from profit audits. We decompose the gross targeting value of data into the net reduction in the tax gap and the implementation cost. This yields a simple adoption criterion: the targeting value must exceed the implementation cost together with the cost of establishing the information system.
Naively conditioning
on manipulable data can perform worse than ignoring the data
altogether.

\subsection*{Illustrative example}

\begin{figure}[H]
\centering
\caption{Data-contingent enforcement}
\label{fig:example_}

\begin{tikzpicture}[line cap=round,line join=round]

\draw[-stealth, line width=1pt] (0,0) -- node[above,rotate=90]{\tiny Tax paid} (0,4)
      node[left,yshift=-.2cm]{\footnotesize $T$};

\draw[-stealth, line width=1pt] (0,0) -- node[below]{\tiny Profit} (6,0)
      node[below,xshift=-.2cm]{\footnotesize $y$};

\draw[dashed] (0,0) -- (6,4);
\node at (5.25,3.5) [below,rotate=37] {\footnotesize $ty$};

\draw[dash pattern={on 10pt off 10pt},
   line width=2pt,
    color=lightblueprint
] (.3,.2) -- (4,2.666) -- (6,2.666)
node[black,right]{\tiny Ordinary activity};

\draw[
    line width=2pt,
    solid,
    color=lightblueprint
] (0,0) -- (2,1.333) -- (6,1.333)
node[black,right]{\tiny Eligible activity};

\draw[dotted] (2,1.333) -- (2,0)
node[below]{\footnotesize $y'_e$};

\draw[dotted] (4,2.666) -- (4,0)
node[below]{\footnotesize $y'_o$};

\end{tikzpicture}
\end{figure}
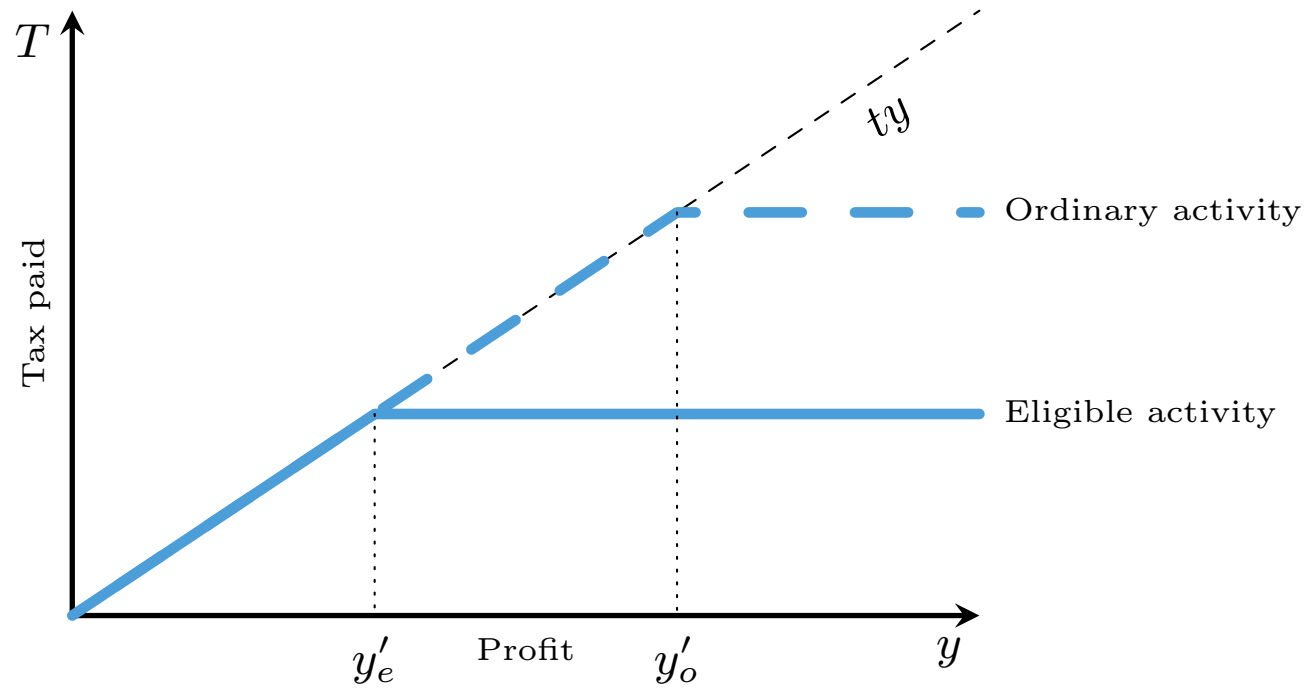

A simple example illustrates the fragility of data-contingent enforcement when classification data are manipulable. Suppose firms self-classify an activity as either \textit{eligible} for special tax treatment (e.g., R\&D expenses) or \textit{ordinary} (e.g., other expenses). Based on this classification, the tax authority sets a lower profit-audit cutoff, $y'_e$, for eligible activities and a higher cutoff, $y'_o$, for ordinary activities. As shown in Figure~\ref{fig:example_}, this generates two effective tax schedules: the blue solid line for eligible activities and the blue dashed line for ordinary activities.

If classification were taken at face value, a firm with an ordinary activity could obtain the more favorable treatment by reporting its activity as eligible and simultaneously reducing its declared tax base to $y'_e$. The firm, therefore, manipulates both dimensions of its report: classification data determine which enforcement schedule it faces, while the profit declaration determines its position on that schedule. Such joint deviations are the central incentive problem in our model.

Credible enforcement of different tax schedules for eligible and ordinary activities requires data verification. 
With two classes, 
the higher cutoff, $y'_o$, defines the global verification frontier: every report below it must be verified either through a profit audit or through verification of the reported classification. Ordinary activities are profit-audited up to $y'_o$; eligible activities are profit-audited up to $y'_e$, while their classification data are verified for $y \in (y'_e,y'_o]$ to induce truthful reporting.\footnote{Since firms with true profits above \(y'_e\) bunch at this cutoff, data verification applies to all such firms, including those whose true profits exceed the global frontier--potentially binding substantial resources.}

\bigskip

The paper is organized as follows. The model is set up in Section \ref{sec:model}. Section \ref{sec:non-strategie} studies non-strategic data reporting, and Section \ref{sec:strategic} investigates strategic data reporting. Section \ref{sec:taxgap} studies the value of data and compares the tax gap in alternative information environments. Section \ref{sec:lit} discusses related literature, and Section \ref{sec:conclu} concludes.
Proofs and omitted results are in Appendices \ref{app:proofs}--\ref{app:taxgap}; extensive supplementary material--on MVPF, comparative statics, analytical examples, and a quantitative illustration--is delegated to Appendices \ref{app:mvpf}--\ref{app:quantitative}.

\section{Model \label{sec:model}}

\paragraph{Setup.}
The economy consists of a unit mass of firms maximizing after-tax profits net of expected penalties for misreporting. A tax authority (hereafter, the IRS) designs an audit policy to minimize the expected tax gap, subject to a budget constraint.

\paragraph{Firms.}
Each firm has a private type $(y,d)$, where $y$ denotes taxable profit and $d$ denotes correlated self-reported data (e.g., turnover, payroll, asset holdings, expenses within a given category) that are used for improved targeting of tax enforcement.\footnote{For multidimensional data, $d$ can be interpreted as a composite risk index used by the authority. For example, the IRS uses its Discriminant Inventory Function (DIF) to score returns for audit targeting.}

Types are drawn from a joint cumulative distribution function  $H$ with a differentiable density function $h$, supported on
$S:=[\underline y,\overline y]\times[\underline d,\overline d]$.
Marginal distributions are denoted by $(H_Y,h_Y)$ and $(H_D,h_D)$, and conditional distributions by $(H_{Y|D},h_{Y|D})$.\footnote{The analysis extends to unbounded support, e.g., $S=[0,\infty)^2$, under the corresponding integrability and tail conditions, ensuring that expected tax liabilities and audit expenditures remain finite.}

\paragraph{Reports.}
After observing its realized type $(y,d)$, each firm simultaneously chooses a profit declaration $x$ and a data declaration $c$ that it reports to the IRS.
The declared profit determines the firm’s statutory tax liability $T(x)$, where $T$ defines an exogenous, increasing tax schedule.
A reporting strategy maps each type $(y,d)$ into declarations $(x,c)$, i.e., $x:=x(y,d)$ and $c:=c(y,d)$.

\paragraph{Audits.}
The IRS commits to an audit policy $A$ that firms can observe before submitting reports $(x,c)$:
\[
A:S\to [0,1]^2, \quad A(x,c):=(\alpha_Y(x,c),\alpha_D(x,c)),
\]
where $\alpha_Y(x,c)$ is the probability of a profit audit and $\alpha_D(x,c)$ is the probability of a data audit for $(x,c)$.

Audits impose costs.
A profit audit entails a cost $\rho_Y$, while a data audit entails a lower cost $\rho_D<\rho_Y$.
Data audits are cheaper because they can rely on partially automated cross-checks with third-party information, whereas profit audits require more detailed verification of accounting information.
Costless data verification, $\rho_D=0$, provides a natural benchmark in which classification data can be verified without using enforcement resources (Section \ref{sec:non-strategie}). We depart from this benchmark by allowing data verification to be arbitrarily cheap but costly, $\rho_D>0$ (Section \ref{sec:strategic}).

Given reporting strategies $(x(y,d),c(y,d))$, expected audit expenditure is
\[
\rho
=
\rho_Y \int \alpha_Y(x(y,d),c(y,d))\,\mathrm dH(y,d)
+
\rho_D \int \alpha_D(x(y,d),c(y,d))\,\mathrm dH(y,d).
\]

We assume that audits perfectly detect misreporting in the dimension being verified. When a data mismatch is detected, the IRS escalates the case to a profit audit, enabling penalties proportional to evaded tax.\footnote{A detected data mismatch triggers a profit audit at additional cost $\rho_Y$; because data reports are truthful in equilibrium, these audit escalation costs arise only off path.}$^{,}$\footnote{\cite{heyes2002theory} presents a theory of two-stage audit where a cursory first-stage audit result may trigger a detailed second stage-audit in the context of emission control.}$^{,}$\footnote{The analysis extends directly to imperfect verification. If a profit audit detects misreporting with probability $\lambda_Y\in(0,1]$ and a data audit detects a discrepancy with probability $\lambda_D\in(0,1]$, effective audit intensities are $\lambda_Y\alpha_Y$ and $\lambda_Y\lambda_D\alpha_D$, respectively.}

\paragraph{Penalties.} If a firm of type $(y,d)$ reports $(x,c)$, its expected penalty is
\[
\Phi(x,c\mid y,d,A)
=
\pi\Bigl(
\alpha_Y(x,c)
+
\alpha_D(x,c)\mathbf 1_{\{c \neq d\}}
\Bigr)
(T(y)-T(x)),
\]
where $\pi>1$ is the marginal penalty per unit of evaded tax.\footnote{We restrict attention without loss to audit policies that do not assign profit and data audits to the same report, so that $\alpha_Y(x,c)\alpha_D(x,c)=0$. A profit audit already verifies the firm's tax liability, making separate verification of its classification data redundant.}
As in \citet{Yitzhaki1974}, the penalty is proportional to the evaded tax, so incentives scale with the extent of profit underreporting. 
Moreover, following \cite{sanchez1993hierarchical}, we assume that firms make truthful reports unless strategic reporting strictly increases expected payoffs. 
Under this standard assumption, firms either report both dimensions truthfully, misreport only profits, or misreport both profits and data; misreporting data alone does not occur.

\paragraph{A firm’s problem.}
Each firm chooses a reporting strategy $(x,c)=(x(y,d),c(y,d))$ to maximize expected net profits after taxes:
\begin{equation}
\Pi(y,d)
=
\max_{x,c}
\left\{
y-T(x)-\Phi(x,c\mid y,d,A)
\right\}.
\label{firm-problem}
\end{equation}

\paragraph{The IRS' problem.}
The IRS chooses an audit policy $A(x,c)$ to minimize the expected tax gap subject to a budget constraint:
\begin{equation}
G(B)
=
\min_{\alpha_Y,\alpha_D}
\int\bigl(T(y)-T(x(y,d))\bigr)\,\mathrm dH(y,d)
\quad\text{s.t.}\quad
\rho\le B.
\label{IRS-problem}
\end{equation}
We assume $B<\rho_Y/\pi(<\rho_Y)$, so full deterrence via profit audits is infeasible.

\paragraph{Timing.}
Anticipating firms' reporting
responses, the IRS commits to $A$. Firms observe their private types $(y,d)$ and submit reports
$(x,c)$ optimally given $A$. The IRS conducts audits according to $A$ and
enforces taxes and penalties.\footnote{Equivalently, the model can be formulated as an extensive game with incomplete information using the Perfect Bayesian equilibrium as the solution concept. Because the IRS commits \textit{ex ante} to the complete audit policy $A(x,c)$, posterior beliefs following reports do not affect subsequent choices; firm and IRS optimality are fully characterized by \eqref{firm-problem} and \eqref{IRS-problem}.}

\section{Benchmark--Truthful data reporting}
\label{sec:non-strategie}

Consider as a benchmark the case where data are truthfully reported.
The IRS can then condition enforcement directly on $d$ and optimally choose a budget $B(d)$ allocated to audit class $d$, with $B=\int B(d) d H_D(d)$.

In a given class $d$, the optimal audit policy can be represented by a cutoff
$y'(d)$, with profit reports below $y'(d)$ audited at the minimal deterrence
probability $\alpha_Y=1/\pi$.\footnote{Otherwise, firms bunch at the lowest
profit level that is not deterred. The cutoff may be interior or lie at the
boundary, corresponding to full enforcement or exemption from profit audits.}
The within-class budget constraint is
\begin{equation}
B(d)
=
\frac{\rho_Y}{\pi}
H_{Y|D}\!\left(y'(d)\mid d\right),
\end{equation}
and the resulting expected tax gap is
\begin{equation}
\label{eq:dGdB_gamma}
G(d)
=
\int_{y'(d)}^{\bar y}
\bigl(T(y)-T(y'(d))\bigr)\,
dH_{Y|D}(y\mid d).
\end{equation}

Differentiating the tax gap with respect to the audit budget yields
\[
\frac{dG(d)}{dB(d)}
=
-
\frac{\pi}{\rho_Y}\,
T'(y'(d))
\gamma(y'(d)\mid d),
\]
where
\[
\gamma(y\mid d)
:=
\frac{1-H_{Y|D}(y\mid d)}
     {h_{Y|D}(y\mid d)}
\]
denotes the inverse hazard rate of profits conditional on data $d$.
This expression gives an \emph{enforcement index}, familiar from optimal-audit models with heterogeneous taxpayer classes \citep{sanchez1993hierarchical,macho2006optimal},
\begin{equation}
\label{eq:enf-ind}
I(y\mid d)
:=
T'(y)\gamma(y\mid d).
\end{equation}
Up to the common factor $\pi/\rho_Y$, the index measures the
reduction in the tax gap from extending audit coverage within class $d$.
It combines the marginal tax liability $T'(y)$ with the \textit{information rents}
generated by private information, summarized by $\gamma(y\mid d)$; Appendix~\ref{app:proofs}
provides the derivation and incentive-theoretic interpretation.

The enforcement index determines how profit audits are allocated. For interior cutoffs, optimal allocation equalizes
marginal enforcement gains across data classes:
\begin{equation}
\label{eq:budget_class}
I\bigl(y'(d)\mid d\bigr)=\kappa,
\end{equation}
where $\kappa>0$ is the shadow value of the audit budget $B$.
Thus, the enforcement index summarizes in which classes additional profit-audit coverage is most valuable.

The slope of the index determines the margin of adjustment.
As depicted in Figure~\ref{fig:optimal_audit_policy}, when
$I(y\mid d)$ is decreasing in $y$, marginal returns decline as audit coverage
expands and enforcement adjusts on the intensive margin through class-specific
cutoffs satisfying $I(y'(d)\mid d)=\kappa$ [Panel (a)].
When $I(y\mid d)$ is non-decreasing, the interior solution satisfying
$I(y'(d)\mid d)=\kappa$ may not be optimal; instead, the IRS chooses between the
boundary cutoffs.\footnote{When the index is increasing, the corner allocation
is determined by the average enforcement index
\[
\bar I(d)
:=
\int_{\underline y}^{\bar y}
I(y\mid d)\,dH_{Y|D}(y\mid d).
\]
A class is fully enforced if $\bar I(d)>\kappa$ and exempt from profit audits
if $\bar I(d)<\kappa$; at equality, the IRS is indifferent. With strategic
data, the analogous condition depends on the endogenous verification frontier;
see Section~\ref{sec:strategic}.}
Optimal enforcement then operates on the extensive margin, with some classes
fully enforced, $y'(d)=\bar y$, and others exempt from profit audits,
$y'(d)=\underline y$ [Panel (b)].\footnote{The standard audit literature focuses on the case with declining marginal returns to enforcement and the familiar cutoff policy \citep{scotchmer1987audit,sanchez1993hierarchical,macho2002auditing,Kuchumova2017}.}


\begin{figure}[!ht]
\centering

\begin{subfigure}{\textwidth}
\centering

\begin{tikzpicture}[
    scale=0.9,
    every node/.style={transform shape},
    line cap=round,
    line join=round
]

\draw[-stealth, line width=1pt] (0,0) -- (6.4,0)
    node[right]{\footnotesize $y$};
\draw[-stealth, line width=1pt] (0,0) -- (0,4.8)
    node[above]{\footnotesize $I(y\mid d)$};

\draw[dashed, line width=1pt] (0,2.1) -- (6.0,2.1);
\node[left] at (0,2.1) {\footnotesize $\kappa$};

\draw[line width=1.6pt, color=lightblueprint]
    (0.3,4.2) .. controls (2.2,3.4) and (4.2,1.7) .. (6.0,0.7);

\draw[line width=1.6pt, dashed, color=lightblueprint]
    (0.3,3.5) .. controls (2.0,2.8) and (4.0,1.3) .. (6.0,0.45);

\coordinate (cutL) at (2.85,2.1);
\coordinate (cutH) at (3.85,2.1);

\draw[dotted, line width=1pt] (cutL) -- (2.85,0);
\draw[dotted, line width=1pt] (cutH) -- (3.85,0);

\node[below] at (2.85,0) {\footnotesize $y'(d_L)$};
\node[below] at (3.85,0) {\footnotesize $y'(d_H)$};

\node[color=lightblueprint, right] at (5.2,1.15)
    {\footnotesize $I(y\mid d_H)$};
\node[color=lightblueprint, right] at (4.9,0.45)
    {\footnotesize $I(y\mid d_L)$};

\node[align=center] at (3.2,4.45)
    {\footnotesize Interior audit cutoffs};

\end{tikzpicture}

\caption{Intensive margin}
\end{subfigure}
\hfill

\begin{subfigure}{\textwidth}
\centering

\begin{tikzpicture}[
    scale=0.9,
    every node/.style={transform shape},
    line cap=round,
    line join=round
]

\draw[-stealth, line width=1pt] (0,0) -- (6.4,0)
    node[right]{\footnotesize $y$};
\draw[-stealth, line width=1pt] (0,0) -- (0,4.8)
    node[above]{\footnotesize $I(y\mid d)$};

\coordinate (ymin) at (0.5,0);
\coordinate (ymax) at (5.5,0);

\node[below] at (ymin) {\footnotesize $\underline y$};
\node[below] at (ymax) {\footnotesize $\bar y$};

\draw[dashed, line width=1pt] (0,2.1) -- (6.0,2.1);
\node[left] at (0,2.1) {\footnotesize $\kappa$};

\draw[line width=1.6pt, color=lightblueprint]
    (0.5,0.55)
    .. controls (2.0,0.85) and (3.6,2.15)
    .. (5.5,4.05);

\draw[line width=1.6pt, dashed, color=lightblueprint]
    (0.5,0.25)
    .. controls (2.0,0.55) and (4.0,1.9)
    .. (5.5,3.15);

\fill (3.45,2.1) circle (1.6pt);
\fill (4.25,2.1) circle (1.6pt);

\draw[dotted, line width=1pt] (3.45,2.1) -- (3.45,0);
\draw[dotted, line width=1pt] (4.25,2.1) -- (4.25,0);

\node[color=lightblueprint, right] at (4.75,3.45)
    {\footnotesize $I(y\mid d_H)$};

\node[color=lightblueprint, right] at (4.75,2.55)
    {\footnotesize $I(y\mid d_L)$};

\fill (0.5,0) circle (1.7pt);
\fill (5.5,0) circle (1.7pt);

\node[align=center, below] at (1.25,-0.32)
    {\scriptsize $y'(d_L)=\underline y$\\[-1pt]
     \scriptsize no enforcement};

\node[align=center, below] at (4.8,-0.32)
    {\scriptsize $y'(d_H)=\bar y$\\[-1pt]
     \scriptsize fully enforced};

\end{tikzpicture}

\caption{Extensive margin}
\end{subfigure}

\caption{\textit{The enforcement index and optimal audit coverage.}}
\label{fig:optimal_audit_policy}
\end{figure}

Together, the preceding analysis explains how auxiliary data facilitate audit targeting by identifying classes in which resources have the highest
marginal value. When data become strategic, however, differentiated
enforcement creates incentives for firms to misclassify their information into more leniently
enforced classes. The upcoming section specifies how data verification resolves this incentive problem while preserving differentiated audit targeting.

\section{Equilibrium--Strategic data reporting}
\label{sec:strategic}

We now allow firms to manipulate the data used for targeting optimal enforcement.
As in the truthful-data benchmark, profit deterrence is characterized by
class-specific cutoffs $y'(d)$: firms report profits truthfully below the
cutoff and bunch at $y'(d)$ above it.

Strategic data reporting introduces a cross-class incentive constraint.
When cutoffs differ across classes, a firm assigned to a more strongly
enforced class can jointly underreport profits and claim membership in a
more leniently enforced class (see the example offered in the introduction). Preventing such deviations requires
verification of the reported data wherever a lower-cutoff class would
otherwise provide an escape from enforcement.

Let
\[
\tilde y:=\inf_d y'(d),
\qquad
\hat y:=\sup_d y'(d),
\]
and call $\hat y$ the (global) \emph{verification frontier}.

\begin{lemma}[Global verification frontier]
\label{lem:global_frontier}
Let $\tilde y<\hat y$.
Every report with $x<\hat y$ must be verified
in equilibrium through either a profit audit or
data verification. Without loss of generality, the optimal hybrid audit policy
can be represented as
\[
\alpha_Y(x,d)
=
\frac{1}{\pi}\mathbf 1_{\{x<y'(d)\}},
\qquad
\alpha_D(x,d)
=
\frac{1}{\pi}
\mathbf 1_{\{y'(d)\le x<\hat y\}}.
\]
\end{lemma}

\begin{proof}
By definition, reports with profits below $\tilde y$ are profit-audited in
every data class. Now suppose that some report
$x\in[\tilde y,\hat y)$ receives neither form of verification in a given data
class $c$. A firm with profit $y>x$ belonging to a class $d$ with cutoff
$y'(d)>x$ could then jointly report data $c$ and profit level $x$ and escape
enforcement. Hence, every report below the global frontier must be covered by
either a profit audit or data verification.
Data audits need not extend beyond $\hat y$: any firm $(y,d)$
can report its true data $d$ and profit
$\min\{y,y'(d)\}\le\hat y$ without suffering penalties,
and thus cannot gain from reporting $x>\hat y$,
which would carry a higher tax. A profit audit directly deters profit underreporting,
while data verification deters joint misreporting
by triggering profit verification when a data
discrepancy is detected. Under the maintained
penalty structure, both instruments provide
deterrence at probability $1/\pi$, yielding
the stated representation.
\end{proof}

For a given cutoff schedule, the resulting hybrid policy assigns profit
audits to reports
\[
x<y'(d)
\]
and data verification to
\[
y'(d)\le x<\hat y.
\]
Consequently, the audit technologies jointly cover all reports below the global frontier.

Note also that firms with true profits above their class-specific profit-audit cutoff bunch at that cutoff. Since the bunching report is subject to data verification when \(y'(d)<\hat y\), all such firms undergo data verification, including those with true profits above the global frontier. Therefore, the cost of data verification depends on the mass of firms above the class-specific cutoff, rather than merely those with profits between the cutoff and the frontier. However, whenever \(y'(d)=\hat y\), the bunching report lies outside the verification region, and no data verification is thus required in that class.

\subsection{Intensive-margin enforcement}

Consider first the case in which the enforcement index
$I(y\mid d)=T'(y)\gamma(y\mid d)$ is strictly decreasing in profits
(e.g., a linear tax schedule $T(y)=\tau y$ and Weibull profits
with $k>1$).
Given a fixed frontier $\hat y$, expected enforcement expenditure is
\begin{equation}
\rho
=
\frac{1}{\pi}
\int
\left[
\rho_Y H_{Y|D}(y'(d)\mid d)
+
\rho_D
\bigl(1-H_{Y|D}(y'(d)\mid d)\bigr)
\mathbf{1}_{\{y'(d)<\hat y\}}
\right]\,dH_D(d).
\label{eq:rho_strategic_rewrite_main}
\end{equation}
The first component captures expenditure on profit audits.
The second captures data-verification expenditure in classes whose
profit-audit cutoff lies below the global frontier.
As firms with profits above their
class-specific cutoff bunch at that cutoff, 
in these classes,
data verification applies to the entire upper tail of the profit distribution. When the cutoff coincides with the global frontier,
no data verification is required.

For classes with profit-audit cutoffs below the global frontier,
the optimal allocation follows the same marginal condition as
in the truthful-data benchmark:
\begin{equation}
I\bigl(y'(d)\mid d\bigr)=\sigma,
\label{eq:strategic_index_main}
\end{equation}
whenever the cutoff is interior.
The difference is that expanding profit audits now replaces
data verification, reducing their marginal cost from
$\rho_Y$ to $\rho_Y-\rho_D$, as reflected in the shadow value of profit-audit coverage
$\sigma=\lambda(\rho_Y-\rho_D)/\pi$.\footnote{
The global verification frontier does not require data verification
to be cheaper than profit auditing. If $\rho_D\geq\rho_Y$, however,
profit audits weakly dominate data verification below the frontier,
and the hybrid enforcement policy collapses to profit auditing.}

However, satisfying this marginal condition need not be globally optimal:
extending profit audits to the global frontier eliminates
data-verification expenditure altogether.
The following proposition characterizes this additional choice.

\begin{proposition}[Intensive-margin enforcement]
\label{prop:strategic_intensive}
Suppose that $I(y\mid d)$ is continuous and strictly decreasing
in $y$ for every data class $d$.
For a given global frontier $\hat y$, let $q<\hat y$ be the
class-specific cutoff implied by the local enforcement index $I(y\mid d)$.

Extending profit audits to the global frontier is preferred
to stopping at $q$ if and only if
\[
\int_q^{\hat y}
\bigl[\sigma-I(y\mid d)\bigr]
h_{Y|D}(y\mid d)\,dy
\le
\frac{\lambda\rho_D}{\pi}
\bigl[1-H_{Y|D}(\hat y\mid d)\bigr].
\]
\end{proposition}

Proposition~\ref{prop:strategic_intensive} characterizes the
tradeoff between differentiated audit targeting and the cost of
preserving truthful data reporting.
The left-hand side measures the loss from extending profit audits
beyond the cutoff favored by marginal enforcement returns.
The right-hand side captures the resulting savings in
data-verification expenditure.
As in the truthful-data benchmark, diminishing marginal returns
favor differentiated, class-specific profit-audit cutoffs.
Under strategic data reporting, however, the savings from
eliminating data verification may outweigh these gains.
Consequently, the optimal policy may combine classes with
lower profit-audit cutoffs and data verification with classes
in which profit audits extend all the way to the global frontier.

\begin{proposition}[Global verification frontier]
\label{prop:optimal_frontier}
Suppose that $I(y\mid d)$ is continuous and strictly
decreasing in $y$ for every data class $d$.
Let $\mathcal D=\{d:y'(d)<\hat y\}$.
If the optimal global verification frontier $\hat y$
is interior and the set $\mathcal D$ remains unchanged
under sufficiently small perturbations of $\hat y$,
then\[
\int_{\mathcal D^c}
\left[
I(\hat y\mid d)-\frac{\lambda\rho_Y}{\pi}
\right]
h_{Y|D}(\hat y\mid d)\,dH_D(d)=0,
\]
where $\lambda>0$ is the multiplier on the
enforcement budget constraint.
\end{proposition}

The condition balances the marginal gains from extending profit audits against their marginal costs, aggregated over classes audited up to the global frontier. For classes with lower profit-audit cutoffs, the frontier does not directly affect equilibrium reporting or verification expenditure. Thus, Proposition \ref{prop:optimal_frontier} demonstrates that the optimal frontier is determined by the classes that rely exclusively on profit audits.

\begin{figure}[htbp]
    \centering
    \includegraphics[width=0.99\textwidth]
    {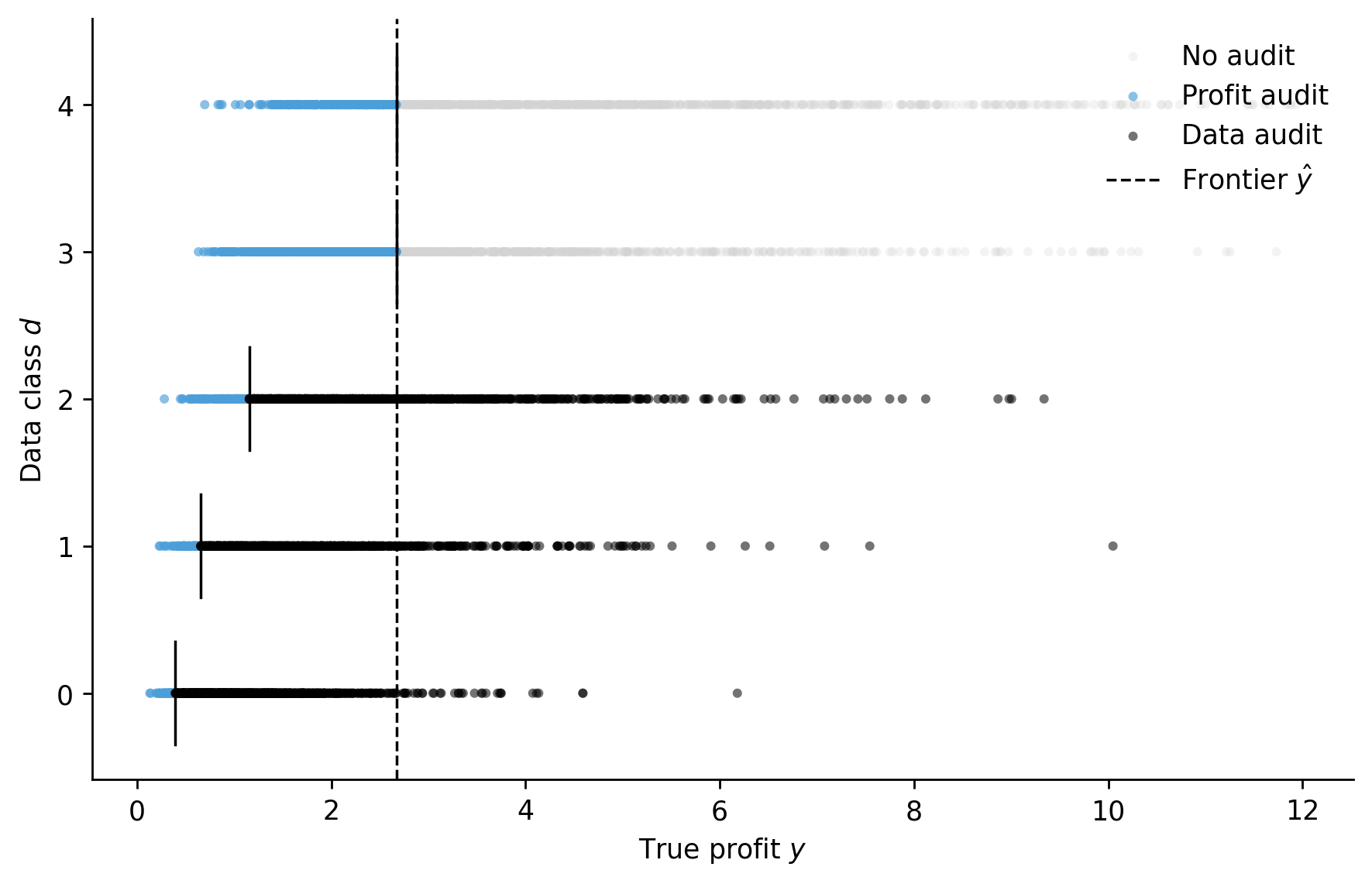}
    \caption{\textit{Cross-sectional audit allocation by true profit.}}
    \label{fig:cross_sec}
\end{figure}

Figure~\ref{fig:cross_sec} describes the allocation of audits
across data classes and true profits.
Profit audits can be concentrated in classes with high enforcement
returns, while data verification sustains this differentiation
by preventing firms from selecting into more leniently enforced
classes.
The two instruments also reach different parts of the true-profit
distribution: profit audits tend to target lower-profit firms,
whereas data verification reaches higher-profit firms whose
reports bunch at class-specific audit cutoffs.
As demonstrated by \citet{macho2002auditing}, access to data can
reverse the regressive bias in tax enforcement.
The figure illustrates how hybrid enforcement can mitigate this
bias by extending enforcement to higher-profit firms.

\label{rem:hybrid_implementation}
Table~\ref{tab:hybrid} further exemplifies the resulting policy for three
classes with
\[
y'(d_1)<y'(d_2)<y'(d_3)=\hat y.
\]
Within each class, profit and data audits partition the verified
region of reported profits: profit audits apply below the
class-specific cutoff and data verification between that cutoff
and the global frontier.
Across classes, firms reporting the same profit may therefore
face different audit instruments.
The class with the highest cutoff relies exclusively on profit
audits, while the remaining classes combine both instruments.

\begin{table}[htbp]
\centering
\begin{tabular}{llll}
\toprule
Reported profit $x$ & Class $d_1$ & Class $d_2$ & Class $d_3$ \\
\midrule
$x < y'(d_1)$
    & Profit audit & Profit audit & Profit audit \\
$x \in [y'(d_1),y'(d_2))$
    & Data audit & Profit audit & Profit audit \\
$x \in [y'(d_2),\hat y)$
    & Data audit & Data audit & Profit audit \\
$x\geq\hat y$
    & No audit & No audit & No audit \\
\bottomrule
\end{tabular}
\caption{Cross-sectional audit allocation by reported profit.}
\label{tab:hybrid}
\end{table}

\begin{remark}[Implications for bunching patterns]
The policy also has a distinctive implication for reporting behavior:
profit reports bunch at class-specific audit cutoffs, whereas
correctly implemented data verification eliminates incentives
to manipulate classification data.
\end{remark}

Empirical studies document bunching of declared amounts around
tax and enforcement thresholds, with excess mass concentrated
on the side favored by the incentives created by the threshold
\citep{saez2010taxpayers,chetty2011adjustment,kleven2013using}.
Our model similarly predicts strategic bunching in profit
declarations at the endogenous audit cutoff, on the side that
avoids profit auditing.
Bunching on the audited side would not be explained by these
strategic incentives alone but may reflect optimization frictions.
By contrast, bunching or manipulation in classification data,
as documented by \citet{almunia2018under} and
\citet{chen2021notching}, may be consistent with insufficient
data verification.

\subsection{Extensive-margin enforcement}

The structure of the hybrid policy changes when the enforcement
index is increasing in profits (e.g., truncated Pareto profits
under a linear tax schedule, or a sufficiently progressive tax
schedule), favoring boundary cutoffs.
With strategic data reporting, however, the choice is between
profit auditing up to the global verification frontier and
relying entirely on data verification within a class.

For this comparison, define the frontier-dependent average
enforcement index
\begin{equation}
\bar I(d;\hat y)
:=
\frac{
\displaystyle
\int_{\underline y}^{\hat y}
I(y\mid d)\,dH_{Y|D}(y\mid d)
}{
H_{Y|D}(\hat y\mid d)
}.
\label{eq:average_enforcement_index_frontier}
\end{equation}
This index measures the average return to profit auditing
over the region below the global verification frontier.
Classes with sufficiently high average enforcement returns
are profit-audited up to the frontier, whereas classes with
lower returns rely on data verification.
Unlike in the truthful-data benchmark, the choice also depends
on the relative costs of the two enforcement technologies.

\begin{proposition}[Extensive-margin enforcement]
\label{prop:strategic_extensive}
Suppose that $I(y\mid d)$ is continuous and strictly increasing
in $y$ for every data class $d$.
For a given global frontier $\hat y$, a class is either
profit audited up to the frontier or instead subject to
data verification.

Profit auditing up to the global frontier is preferred
to data verification if and only if
\[
\bar I(d;\hat y)
\ge
\frac{\lambda}{\pi}
\left[
\rho_Y-
\frac{\rho_D}{H_{Y|D}(\hat y\mid d)}
\right].
\]
\end{proposition}

The global verification frontier and the allocation of
enforcement across classes are determined jointly,
subject to the aggregate budget constraint.
The proposition reveals a distinction between the two
enforcement technologies.
Without profit audits, all firms in a class bunch at the lowest profit report. Their reports are hence subject to data verification, making the resulting verification expenditure independent of the global frontier.
By contrast, profit-audit expenditure also depends on
the fraction of firms below the frontier whose profits are audited.
Consequently, other things equal,
profit auditing becomes more attractive in classes with
a smaller fraction of firms below the global frontier.
Profit audits thus concentrate in higher-profit classes while targeting lower-profit firms within them.

\begin{example}[Extending enforcement across classes or profit levels]
Consider two equally likely classes $d_H,d_L$, with profits
$y\in\{0,1,2\}$ and conditional distributions
\[
\begin{array}{c|ccc}
 & y=0 & y=1 & y=2\\ \hline
d_H & 0.6 & 0.2 & 0.2\\
d_L & 0.8 & 0.2 & 0
\end{array}
\]
Let $T(0)=0$, $T(1)=1$, $T(2)=7$,
$\pi=2$, $\rho_Y=2$, and $\rho_D=1.2$.
Profit reports strictly below
the audit cutoff are audited, while firms with higher
profits bunch at the cutoff.

Suppose initially that $\hat y=1$, with $d_H$
profit-audited up to the frontier and $d_L$
subject to data verification.
The IRS can either extend profit auditing to $d_L$
or raise the frontier to $\hat y=2$, retaining
the original class allocation.
Both expansions cost $0.10$ in aggregate,
but their reductions in the aggregate tax gap are,
respectively,
$\frac{0.2}{2}T(1)=0.1$ and
$\frac{0.2}{2}[T(2)-T(1)]=0.6$.
Thus, extending the frontier yields the greater return.
If instead $T(2)=1.5$, this benefit falls to $0.05$,
reversing the ranking in favor of expanding
audits across classes.
\end{example}

While the shape of the enforcement index determines how profit audits are allocated, Propositions~\ref{prop:strategic_intensive} and \ref{prop:strategic_extensive} reveal a common principle. The verification frontier is driven by the cross-class incentive constraint: whenever taxpayers can manipulate the tax base by misreporting their class, targeted enforcement requires data verification in more leniently enforced classes.

\begin{remark}[Profit-audit exemptions]
\label{rem:exempt_classes}
With strategic data, exemption from profit audits does not imply
the absence of enforcement. An unverified class otherwise creates an
enforcement loophole, because firms can avoid profit-audits through data
misclassification. Exempt classes must, therefore, be data-verified wherever
they provide an escape from profit audit.\footnote{Although the statutory tax schedule is fixed, the optimal audit
policy can generate a \textit{de facto} tax exemption: in classes exempt
from profit audits, firms bunch at the lowest reported profit
$\underline y$, so statutory tax liabilities above $T(\underline y)$ are
not enforced. When $\underline y=0$ and $T(0)=0$, this amounts to zero
effective tax liability.}
\end{remark}

Our characterization identifies both the benefit and the cost of
data-contingent enforcement. Data improve targeting by allowing profit audits
to be directed toward classes with higher enforcement returns. When data are
strategic, however, sustaining this differentiation requires additional
verification. 
The following section investigates this tradeoff more closely.

 \section{Value of data}\label{sec:taxgap}
 
We characterize the value of introducing self-reported data into
enforcement by comparing the expected tax gap across three enforcement
regimes: $G^A$ when the IRS ignores data, $G^B$ when data are truthful, and
$G^C$ when data are strategic and optimally verified. The corresponding tax gaps are
\[
G^A
=
\int_{\hat y_A}^{\bar y}
\bigl(T(y)-T(\hat y_A)\bigr)dH_Y(y),
\]
and, for $j\in\{B,C\}$,
\[
G^j
=
\int\!\!\int_{y'_j(d)}^{\bar y}
\bigl(T(y)-T(y'_j(d))\bigr)
dH_{Y|D}(y\mid d)dH_D(d),
\]
where the cutoffs $y'_B(d)$ and $y'_C(d)$ are characterized in
Sections~\ref{sec:non-strategie} and \ref{sec:strategic}, respectively.
Formal derivations are provided in Appendix~\ref{app:taxgap}.

Because the IRS can always ignore classification data, and truthful data
relax the incentive constraints imposed by strategic reporting,
\[
G^B\leq G^C\leq G^A.
\]
The difference between $G^B$ and $G^A$ identifies the \emph{targeting value} of data:
\[
V_T:=G^A-G^B.
\]
It measures the reduction in the tax gap attainable when classification data
can be used without cost to direct profit audits toward classes with the highest
enforcement returns.

When data are strategic, part of this targeting value must be sacrificed to
make differentiated enforcement incentive compatible. Under the optimal
hybrid policy, the enforcement budget is divided between profit audits,
\[
B_Y
=
\frac{\rho_Y}{\pi}
\int
H_{Y|D}(y'_C(d)\mid d)
dH_D(d),
\]
and data verification,
\[
B_D
=
\frac{\rho_D}{\pi}
\int
\bigl[1-H_{Y|D}(y'_C(d)\mid d)\bigr]
\mathbf{1}_{\{y'_C(d)<\hat y_C\}}
\,dH_D(d),
\]
with $B_Y+B_D=B$. The opportunity cost of this verification for the IRS  is
\[
C_S:=G^C-G^B.
\]
Hence, the net enforcement value of strategic data admits the decomposition
\begin{equation}
\label{eq:value_decomposition}
G^A-G^C
=
\underbrace{G^A-G^B}_{\text{targeting value}}
-
\underbrace{(G^C-G^B).}_{\text{implementation cost}}
\end{equation}
\begin{proposition}[Value of data]\label{prop:value_data}
The tax gaps under truthful data, optimally verified strategic data, and no
data satisfy
\[
G^B\leq G^C\leq G^A.
\]
The gross targeting value of classification data is $G^A-G^B$, while their
net enforcement value under strategic reporting is
\[
G^A-G^C=(G^A-G^B)-(G^C-G^B).
\]
\end{proposition}

\begin{example}[Informative data without enforcement value]
Consider two equally likely data classes, $d_1$ and $d_2$,
with profits $y\in\{0,1,2\}$ distributed as follows:
\[
\begin{array}{c|ccc}
 & y=0 & y=1 & y=2\\ \hline
d_1 & 0.40 & 0.05 & 0.55\\
d_2 & 0.90 & 0.01 & 0.09
\end{array}
\]
Let $T(y)=y$, $\pi=2$, $\rho_Y=2$,
$\rho_D=1.9$, and $B=0.65$.
Reports are restricted to the three profit levels,
and profit reports strictly below the cutoff are audited.

Without data, the optimal uniform cutoff is $1$,
yielding $G^A=0.32$.
With truthful data, the IRS instead chooses
$y'_B(d_1)=2$ and $y'_B(d_2)=0$,
yielding $G^B=0.095$.

When data are strategic, however, every differentiated
cutoff allocation requires data verification and exceeds
the budget. The IRS therefore retains the uniform cutoff
$1$, implying $G^B<G^C=G^A$.
Thus, informative classification data can have no
enforcement value when reporting is strategic.
\end{example}

Suppose, in addition, that introducing classification data into the
enforcement system entails a fixed cost $K\geq0$. This may capture any costs of data administration or IT infrastructure not included in the audit
budget. Data should then be introduced if and only if
\begin{equation}
\label{eq:data_adoption}
\underbrace{G^A-G^B}_{\text{targeting value}}
\geq
\underbrace{G^C-G^B}_{\text{implementation cost}}
+
\underbrace{K.}_{\text{data-introduction cost}}
\end{equation}
Thus, introducing informative data into tax enforcement is not necessarily valuable to the authority: their targeting
gains must be sufficiently large to compensate both for the resources required
to sustain truthful classification and for the cost of introducing the data.

Verification is essential for realizing targeting gains.
If the IRS conditions enforcement on manipulable data
without verifying them, firms select into the most
leniently enforced class.
Writing the resulting pooling cutoff as
$\tilde y=\inf_d y'(d)$, the tax gap becomes
\[
G^D
=
\int_{\tilde y}^{\bar y}
\bigl(T(y)-T(\tilde y)\bigr)dH_Y(y).
\]
For a budget-feasible pooling policy,
$G^D\geq G^A\geq G^C$.
The first inequality follows because budget feasibility
requires $\tilde y\leq\hat y_A$, with strict inequality
whenever $\tilde y<\hat y_A$.
Thus, naive reliance on manipulable data can perform
worse than ignoring them altogether.

\begin{corollary}\label{coro:value_data}
Introducing manipulable data at cost $K$ is optimal if and only if
\[
G^A-G^B\geq (G^C-G^B)+K.
\]
If data are used naively without verification, then $G^D\geq G^A(\geq G^C\geq G^B)$.
\end{corollary}

\begin{figure}[htbp]
\centering
    \includegraphics[width=0.8\textwidth]
    {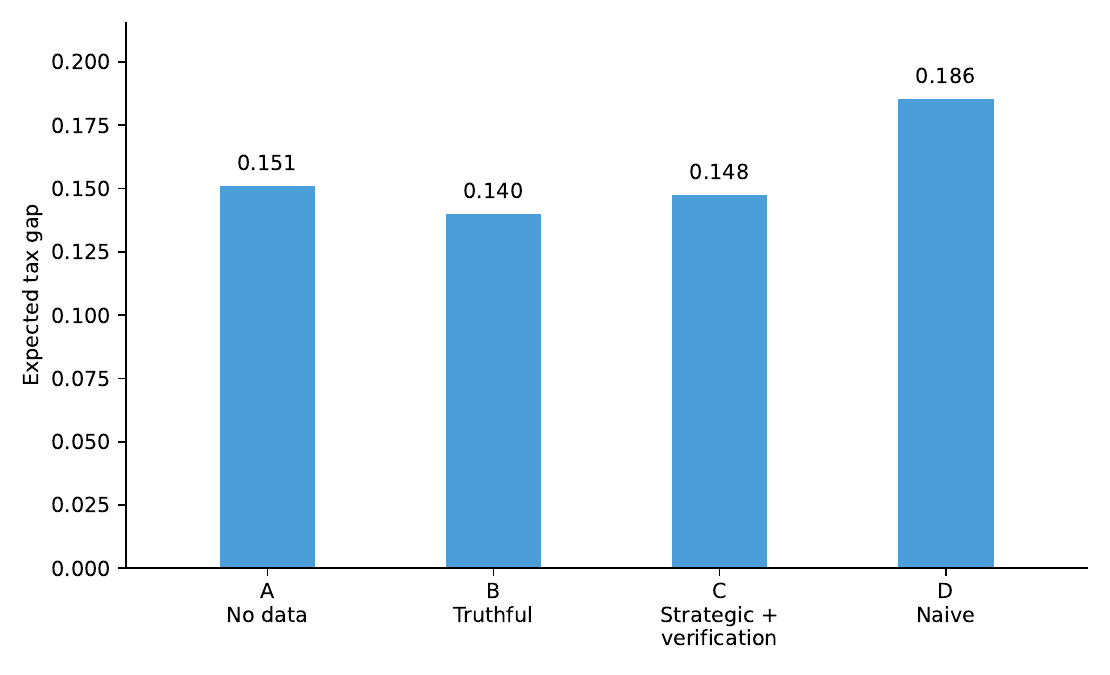}
\caption{\textit{Tax gaps across enforcement regimes.}}
\label{fig:gap_comp}
\end{figure}

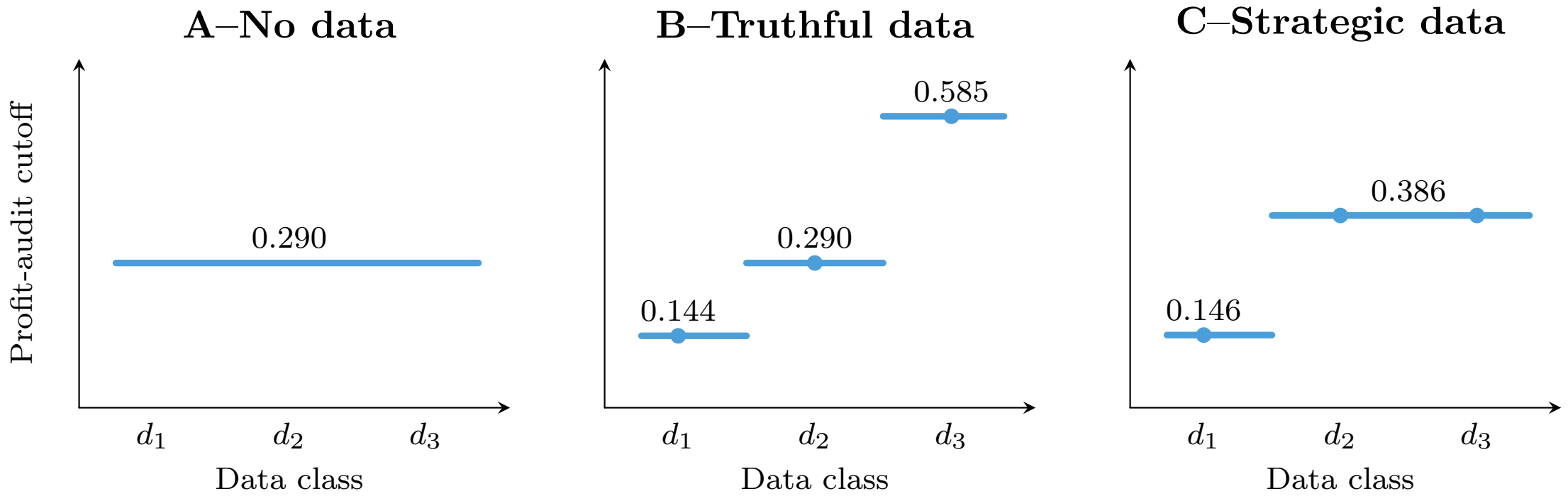
\begin{figure}[htbp]
\centering
\resizebox{\textwidth}{!}{%
\begin{tikzpicture}[line cap=round,line join=round,>=stealth,x=0.95cm,y=4.5cm]
\begin{scope}[shift={(0,0)}]
\node[font=\small\bfseries] at (2,0.77) {A--No data};
\draw[->] (0,0) -- (4.1,0);
\draw[->] (0,0) -- (0,0.70);
\node[rotate=90,font=\scriptsize] at (-0.55,0.35) {Profit-audit cutoff};
\node[below,font=\scriptsize] at (0.7,0) {$d_1$};
\node[below,font=\scriptsize] at (2,0) {$d_2$};
\node[below,font=\scriptsize] at (3.3,0) {$d_3$};
\draw[lightblueprint,line width=1.7pt] (0.35,0.290434) -- (3.8,0.290434);
\node[above,font=\scriptsize] at (2,0.290434) {$0.290$};
\node[below,font=\scriptsize] at (2,-0.09) {Data class};
\end{scope}
\begin{scope}[shift={(5,0)}]
\node[font=\small\bfseries] at (2,0.77) {B--Truthful data};
\draw[->] (0,0) -- (4.1,0);
\draw[->] (0,0) -- (0,0.70);
\node[below,font=\scriptsize] at (0.7,0) {$d_1$};
\node[below,font=\scriptsize] at (2,0) {$d_2$};
\node[below,font=\scriptsize] at (3.3,0) {$d_3$};
\draw[lightblueprint,line width=1.7pt]
(0.35,0.144225) -- (1.35,0.144225)
(1.35,0.290434) -- (2.65,0.290434)
(2.65,0.584862) -- (3.8,0.584862);
\fill[lightblueprint] (0.7,0.144225) circle (2pt);
\fill[lightblueprint] (2,0.290434) circle (2pt);
\fill[lightblueprint] (3.3,0.584862) circle (2pt);
\node[above,font=\scriptsize] at (0.7,0.144225) {$0.144$};
\node[above,font=\scriptsize] at (2,0.290434) {$0.290$};
\node[above,font=\scriptsize] at (3.3,0.584862) {$0.585$};
\node[below,font=\scriptsize] at (2,-0.09) {Data class};
\end{scope}
\begin{scope}[shift={(10,0)}]
\node[font=\small\bfseries] at (2,0.77) {C--Strategic data};
\draw[->] (0,0) -- (4.1,0);
\draw[->] (0,0) -- (0,0.70);
\node[below,font=\scriptsize] at (0.7,0) {$d_1$};
\node[below,font=\scriptsize] at (2,0) {$d_2$};
\node[below,font=\scriptsize] at (3.3,0) {$d_3$};
\draw[lightblueprint,line width=1.7pt]
(0.35,0.145772) -- (1.35,0.145772)
(1.35,0.385790) -- (3.8,0.385790);
\fill[lightblueprint] (0.7,0.145772) circle (2pt);
\fill[lightblueprint] (2,0.385790) circle (2pt);
\fill[lightblueprint] (3.3,0.385790) circle (2pt);
\node[above,font=\scriptsize] at (0.7,0.145772) {$0.146$};
\node[above,font=\scriptsize] at (2.65,0.385790) {$0.386$};
\node[below,font=\scriptsize] at (2,-0.09) {Data class};
\end{scope}
\end{tikzpicture}%
}
\caption{\textit{Profit-audit cutoffs across enforcement regimes.}
}
\label{fig:cutoff_comparison}
\end{figure}

Figure~\ref{fig:gap_comp} describes the tax-gap comparisons in
Proposition~\ref{prop:value_data} and Corollary~\ref{coro:value_data}, based
on the calibrated quantitative illustration in Appendix~\ref{app:quantitative}.
Moving from the no-data benchmark (A) to the truthful-data benchmark (B) reduces the tax gap from $G^A=0.151$ to $G^B=0.140$, yielding a gross targeting value of $0.011$. Under strategic reporting, optimal verification (C) yields $G^C=0.148$, whereas naive reliance on unverified data (D) yields $G^D=0.186$.
At the baseline relative verification cost $\rho_D/\rho_Y=0.03$,
verification hence absorbs about 9.5 percent of the enforcement budget,
and strategic reporting dissipates approximately 67 percent of the
gross targeting value of classification data.
Interestingly, as illustrated in Figure~\ref{fig:cutoff_comparison},
the optimal policy tailored for strategic data reporting features partial
pooling of enforcement cutoffs: two of the three classes share a
profit-audit cutoff, while the third is treated differently.
Thus, verification costs affect not only the value of
data but also the optimal degree of class-by-class differentiation.

\section{Literature}\label{sec:lit}

Our analysis builds on the literature studying how auxiliary information can improve the design of tax enforcement.
\citet{scotchmer1987audit} shows that audit probabilities should condition on taxpayer characteristics and bounds the maximum audit probability by the inverse of penalties.
\citet{sanchez1993hierarchical} establishes the familiar cutoff structure in which evasion is deterred up to an income threshold, above which taxpayers are not audited.
\citet{Macho1997} extends the analysis to multiple income sources:
audit policies differ across sources, while within each source only incomes
below a source-specific cutoff are audited.
\citet{macho2002auditing} further demonstrate how signals correlated with income can be used to target audits and may reverse the standard regressive bias of optimal enforcement policy.
\citet{cremer1996tax}, \citet{aronsson2026optimal}, and
\citet{canta2024welfare} study optimal taxation with tax evasion in the Mirrleesian tradition with multiple taxpayer types.

Our model bridges the cutoff structure of \citet{sanchez1993hierarchical} and the differentiated enforcement of \citet{macho2002auditing}. As in \citet{macho2002auditing}, auxiliary information allows profit enforcement to differ across data classes. Once the information used for targeting can itself be manipulated, however, these class-specific policies can no longer be chosen independently. Cross-class incentive compatibility generates a single global verification frontier, much like in \citet{sanchez1993hierarchical}: every report below this frontier must be verified. Data verification connects differentiated profit enforcement to this common frontier. When marginal returns to profit auditing are decreasing, profit audits extend to class-specific cutoffs, with data verification filling the remaining gap to the frontier. When marginal returns are increasing, the hybrid structure instead operates class by class: some classes are profit-audited up to the frontier, while others are enforced through cheaper data verification.\footnote{More complex audit patterns can also arise in models with additional information or dimensions of heterogeneity; see, for example, \citet{castillo2024tax,zilberman2016audits}.}

\citet{Kuchumova2017} also endogenizes the information available for enforcement by allowing the tax authority to invest in the accuracy of external information reporting.\footnote{\citet{menichini2014sorting}, \citet{Boserup2013}, and \citet{Phillips2014} allow for additional information in auditing but do not focus on firms' reporting incentives.}
The authority must then decide how much of a limited enforcement budget to devote to improving information and how much to audits. More accurate information improves targeting, but its value falls with audit coverage. Our approach instead takes the informativeness of classification data as given but allows taxpayers to manipulate the information used for targeting. Data verification then has a direct implementation role: it prevents taxpayers from selecting into more leniently enforced classes and thereby sustains differentiated profit enforcement. This creates a tradeoff between the targeting gains from classification data and the verification resources required to use them. We characterize the resulting net value of data by comparing optimal enforcement with strategic data to the implementable benchmark in which the authority ignores classification data.

Our treatment of classification data is motivated by a broad empirical literature on behavioral responses to the design of tax policy and enforcement. This literature indicates that taxpayers respond to different policies not only along the conventional margin of income misreporting, but also by changing reported characteristics on which tax rates, deductions, exemptions, and audit treatment are later conditioned \citep{almunia2018under,almunia2022strategic,best2015production,elschner2013special,fisman2004tax,chen2021notching,aghion2023tax,kuchumova2025beyond}.

\section{Conclusion}\label{sec:conclu}

How can tax authorities use detailed taxpayer data without introducing
new opportunities for evasion?
When classification data are potentially manipulable,
we demonstrate that optimal enforcement is organized around a global verification frontier,
with data verification preventing firms from dodging enforcement through misclassification.
The structure of optimal hybrid tax enforcement
depends on whether profit auditing operates on the intensive or extensive margin,
with the costs of data verification restricting differentiation across data classes.
Data are therefore valuable only when
the associated targeting gains exceed the costs of verification and
those of introducing data into enforcement;
without credible verification, reliance on self-classified data
can perform worse than disregarding data altogether.
When successfully implemented, the optimal policy generates bunching in
profits but involves no data bunching, providing an empirical diagnostic for
insufficient data verification.
Taken together, these results provide the first unified account of how
the possibility of manipulating classification data affects optimal tax
enforcement and the resulting patterns of reporting behavior.

\section*{Appendix}

\appendix

\section{Omitted results}
\label{app:proofs}

\subsection{Truthful data reporting}

For completeness, the following analysis derives the enforcement index and information rents directly from the
IRS problem (Appendix \ref{app:index_derivation}) and then provides the mechanism-design interpretation of optimal audit policy (Appendix \ref{app:implementability}).

\subsubsection{Enforcement index and information rents}
\label{app:index_derivation}

Fix a data class $d$ and write $y':=y'(d)$. From
\[
G(d)
=
\int_{y'}^{\bar y}
\bigl(T(y)-T(y')\bigr)\,
dH_{Y|D}(y\mid d),
\]
Differentiating this using Leibniz's rule gives
\[
\frac{dG(d)}{dy'}
=
-\bigl(T(y')-T(y')\bigr)h_{Y|D}(y'\mid d)
-
T'(y')
\int_{y'}^{\bar y}dH_{Y|D}(y\mid d),
\]
and hence
\[
\frac{dG(d)}{dy'}
=
-
T'(y')
\bigl(1-H_{Y|D}(y'\mid d)\bigr).
\]
Increasing the cutoff by $dy'$ raises the reported tax liability of every
firm above the cutoff by approximately $T'(y')dy'$. The mass of these firms
is $1-H_{Y|D}(y'\mid d)$. The boundary term vanishes because a firm at the
cutoff has no tax gap.

The within-class budget satisfies
\[
B(d)
=
\frac{\rho_Y}{\pi}
H_{Y|D}(y'\mid d),
\qquad
\frac{dB(d)}{dy'}
=
\frac{\rho_Y}{\pi}
h_{Y|D}(y'\mid d),
\]
which gives
\[
\frac{dG(d)}{dB(d)}
=
-
\frac{\pi}{\rho_Y}
T'(y')
\frac{1-H_{Y|D}(y'\mid d)}
     {h_{Y|D}(y'\mid d)}
=
-
\frac{\pi}{\rho_Y}
I(y'\mid d).
\]

The same expression gives the information-rent interpretation of the index.
A marginal increase in the cutoff by $dy'$ reduces the reporting rent of
every firm above $y'$ by approximately $T'(y')dy'$. The mass of these
inframarginal firms is $1-H_{Y|D}(y'\mid d)$, while the marginal mass that
must be brought under audit coverage is
$h_{Y|D}(y'\mid d)dy'$. Thus the rent reduction per marginal firm brought
under enforcement is
\[
T'(y')
\frac{1-H_{Y|D}(y'\mid d)}
     {h_{Y|D}(y'\mid d)}
=
T'(y')\gamma(y'\mid d)
=
I(y'\mid d).
\]
The inverse hazard rate therefore summarizes the information rents generated
by incomplete enforcement.

\subsubsection{Mechanism design and optimal enforcement}
\label{app:implementability}

The optimal profit-audit policy can also be derived using standard mechanism-design
arguments. 

Fix a data class $d$. Consider a firm with profit $y$ contemplating
a lower profit report $x<y$. Relative to truthful reporting, underreporting of profit yields
\[
\bigl(1-\pi\alpha_Y(x,d)\bigr)
\bigl[T(y)-T(x)\bigr].
\]
As $T$ is increasing, tax evasion is deterred if and only if
\[
\alpha_Y(x,d)\geq\frac{1}{\pi}.
\]
Because audits are costly, the IRS applies the minimum audit probability $1/\pi$ required
for deterrence on enforced reports and otherwise sets a zero probability. Therefore, in an optimal audit
policy,
\[
\alpha_Y(x,d)\in\left\{0,\frac{1}{\pi}\right\}.
\]

Incentive compatibility further requires the profit-audit probability assigned
to equilibrium reports to be non-increasing in profits. To see this, consider
$y_2>y_1$. The two incentive constraints imply
\begin{align*}
y_1-T(x(y_1,d))
-\Phi(x(y_1,d),d\mid y_1,d,A)
&\geq
y_1-T(x(y_2,d))
-\Phi(x(y_2,d),d\mid y_1,d,A),\\
y_2-T(x(y_2,d))
-\Phi(x(y_2,d),d\mid y_2,d,A)
&\geq
y_2-T(x(y_1,d))
-\Phi(x(y_1,d),d\mid y_2,d,A).
\end{align*}
Adding these constraints and using the definition of $\Phi$ gives
\[
\pi
\left[
\alpha_Y(x(y_1,d),d)
-
\alpha_Y(x(y_2,d),d)
\right]
\left[T(y_2)-T(y_1)\right]
\geq0.
\]
Since taxes are increasing,
\[
\alpha_Y(x(y_1,d),d)
\geq
\alpha_Y(x(y_2,d),d).
\]
Together with $\alpha_Y\in\{0,1/\pi\}$, this implies a cutoff rule: $\alpha_Y(x,d)=1/\pi$ for $x<y'(d)$ and $\alpha_Y(x,d)=0$ for $x\geq y'(d)$.

The envelope theorem applied to the firm's problem yields
\[
\frac{\partial\Pi(y,d)}{\partial y}
=
1-\pi\alpha_Y(x(y,d),d)T'(y).
\]
Integrating from $\underline y$ to $y$,
\[
\Pi(y,d)
=
\Pi(\underline y,d)
+(y-\underline y)
-\pi\int_{\underline y}^{y}
\alpha_Y(x(s,d),d)T'(s)\,ds.
\]
Thus, when profit auditing is positive at $\alpha_Y=1/\pi$ over
$[\underline y,y]$, it reduces the information rent of type $y$ by
$\int_{\underline y}^{y} T'(s)\,ds$.

For a fixed data class, minimizing the tax gap is equivalent to maximizing
expected reported tax payments, since the statutory tax liability
$\int T(y)\,dH_{Y|D}(y\mid d)$ is fixed. From the firm's equilibrium payoff,
\[
T(x(y,d))
=
y-\Pi(y,d)
-\Phi(x(y,d),d\mid y,d,A).
\]
Substituting the envelope representation and taking expectations gives
\begin{align*}
\int_{\underline y}^{\bar y}
T(x(y,d))\,dH_{Y|D}(y\mid d)
={}&
\underline y-\Pi(\underline y,d)
\\
&+
\pi\int_{\underline y}^{\bar y}
\int_{\underline y}^{y}
\alpha_Y(x(s,d),d)T'(s)\,ds\,
dH_{Y|D}(y\mid d)
\\
&-
\int_{\underline y}^{\bar y}
\Phi(x(y,d),d\mid y,d,A)\,
dH_{Y|D}(y\mid d).
\end{align*}
Applying Fubini's theorem,
\begin{align*}
\int_{\underline y}^{\bar y}
T(x(y,d))\,dH_{Y|D}(y\mid d)
={}&
\underline y-\Pi(\underline y,d)
\\
&+
\pi\int_{\underline y}^{\bar y}
\alpha_Y(x(y,d),d)T'(y)
\bigl[1-H_{Y|D}(y\mid d)\bigr]\,dy
\\
&-
\int_{\underline y}^{\bar y}
\Phi(x(y,d),d\mid y,d,A)\,
dH_{Y|D}(y\mid d).
\end{align*}
Under the optimal implementation, expected penalties vanish on path:
profit-audited reports are truthful, while reports involving profit
understatement are not audited. Hence,
\[
\int_{\underline y}^{\bar y}
T(x(y,d))\,dH_{Y|D}(y\mid d)
=
\underline y-\Pi(\underline y,d)
+
\pi\int_{\underline y}^{\bar y}
\alpha_Y(x(y,d),d)
T'(y)\gamma(y\mid d)\,
dH_{Y|D}(y\mid d).
\]

Consequently, apart from the lowest type's payoff, the IRS's objective assigns
profit-audit coverage the mechanism-design weight
\[
I(y\mid d)=T'(y)\gamma(y\mid d).
\]
The term $T'(y)$ captures the marginal tax liability, while
$\gamma(y\mid d)$ reflects the information rents reduced by extending
enforcement at profit $y$. 

In other words, the enforcement index measures the return
to extending profit-audit coverage. Let $\kappa>0$ denote the shadow value
of the aggregate enforcement budget. With a decreasing index, the IRS acts on these
returns pointwise, extending profit-audit coverage until
\[
I(y\mid d)=\kappa.
\]
With an increasing index, an interior class-specific cutoff is not optimal.
The IRS instead chooses between the boundary policies,
\[
y'(d)\in\{\underline y,\bar y\},
\]
so that a class is either fully profit-audited or exempt from profit audits.
The choice is governed by the average enforcement index
\[
\bar I(d)
=
\int_{\underline y}^{\bar y}
I(y\mid d)\,dH_{Y|D}(y\mid d).
\]
A class is fully enforced if $\bar I(d)>\kappa$ and exempt from audits if
$\bar I(d)<\kappa$, with indifference at equality.\footnote{As the distribution of data classes is continuous, every class has
zero mass, and we can abstract from the partial enforcement of a class.}
Thus, the same enforcement index governs both margins: pointwise returns
determine the intensive-margin allocation, whereas their average determines
the extensive-margin allocation across classes.

\subsection{Strategic data reporting}

\begin{proof}[Proof of Proposition~\ref{prop:strategic_intensive}]
Fix the global frontier $\hat y$ and a data class $d$.
For a profit-audit cutoff $q<\hat y$, expected enforcement
expenditure is
\[
C_d(q)
=
\frac{1}{\pi}
\left[
\rho_Y H_{Y|D}(q\mid d)
+
\rho_D\bigl(1-H_{Y|D}(q\mid d)\bigr)
\right].
\]
Increasing $q$ replaces data verification by profit auditing.
The marginal reduction in the tax gap is
\[
T'(q)\bigl[1-H_{Y|D}(q\mid d)\bigr],
\]
while the marginal resource cost is
\[
\frac{\rho_Y-\rho_D}{\pi}h_{Y|D}(q\mid d).
\]
Let $\mathcal L_d(t)=G_d(t)+\lambda C_d(t)$ denote the
class-specific Lagrangian, where $G_d(t)$ is the expected tax gap
and $\lambda$ is the multiplier on the aggregate enforcement budget.
For $t<\hat y$, its derivative is
\[
\frac{d\mathcal L_d(t)}{dt}
=
h_{Y|D}(t\mid d)
\bigl[\sigma-I(t\mid d)\bigr],
\qquad
\sigma=\frac{\lambda(\rho_Y-\rho_D)}{\pi}.
\]
Since $I(\cdot\mid d)$ is strictly decreasing, the
Lagrangian decreases where $I(t\mid d)>\sigma$
and increases where $I(t\mid d)<\sigma$.
Thus, the preferred cutoff below the frontier is
uniquely determined by
\[
I(q\mid d)=\sigma,
\]
if interior, or by the corresponding lower boundary.
If the Lagrangian is decreasing throughout the interval
below the frontier, extending profit audits to the frontier
is optimal.

At the global frontier, data-verification expenditure
is eliminated altogether, so that
\[
C_d(\hat y)
=
\frac{\rho_Y}{\pi}H_{Y|D}(\hat y\mid d).
\]
Since the tax gap is continuous in the cutoff, the Lagrangian has a downward jump at the frontier of size \(\lambda\rho_D[1-H_{Y|D}(\hat y\mid d)]/\pi\).
This may make extending profit audits to $\hat y$ preferable
to the cutoff $q$ implied by the enforcement index.
Comparing these two cutoffs gives
\begin{align*}
\mathcal L_d(\hat y)-\mathcal L_d(q)
&=
-\int_q^{\hat y}
I(y\mid d)h_{Y|D}(y\mid d)\,dy
\\
&\quad+
\frac{\lambda}{\pi}
\left[
\rho_Y H_{Y|D}(\hat y\mid d)
-\rho_D
-(\rho_Y-\rho_D)H_{Y|D}(q\mid d)
\right]
\\
&=
\int_q^{\hat y}
\bigl[\sigma-I(y\mid d)\bigr]
h_{Y|D}(y\mid d)\,dy
\\
&\quad-
\frac{\lambda\rho_D}{\pi}
\bigl[1-H_{Y|D}(\hat y\mid d)\bigr].
\end{align*}
Thus, extending profit audits to the global frontier is
preferred if and only if this difference is nonpositive,
yielding the stated condition.
\end{proof}

\begin{proof}[Proof of Proposition~\ref{prop:optimal_frontier}]
By Proposition~\ref{prop:strategic_intensive},
conditional on the global frontier $\hat y$,
each class either chooses a lower cutoff
$y'(d)<\hat y$ or extends profit audits
to the frontier.

Consider a local change in $\hat y$.
For classes in $\mathcal D$, the optimal
cutoff is determined by
$I(y'(d)\mid d)=\sigma$, or by the
corresponding lower boundary, and is
independent of $\hat y$.
Their contribution to the Lagrangian
therefore remains unchanged.

For classes in $\mathcal D^c$, the
profit-audit cutoff equals $\hat y$
and no data verification is required.
Using the class-specific Lagrangian
defined in the proof of
Proposition~\ref{prop:strategic_intensive},
we obtain
\[
\frac{d\mathcal L_d(\hat y)}{d\hat y}
=
h_{Y|D}(\hat y\mid d)
\left[
\frac{\lambda\rho_Y}{\pi}
-I(\hat y\mid d)
\right].
\]
Since the allocation of classes between
the two regimes is locally unchanged,
the derivative of the aggregate
Lagrangian is obtained by integrating
this expression over $\mathcal D^c$.
At an interior optimum, this derivative
must vanish, yielding the stated condition.
\end{proof}

\begin{proof}[Proof of Proposition~\ref{prop:strategic_extensive}]
Fix the global verification frontier $\hat y$ and a data class $d$.
For any cutoff $q<\hat y$, expected enforcement expenditure is
\[
C_d(q)
=
\frac{1}{\pi}
\left[
\rho_Y H_{Y|D}(q\mid d)
+
\rho_D\bigl(1-H_{Y|D}(q\mid d)\bigr)
\right].
\]
Let
\[
\mathcal L_d(q)=G_d(q)+\lambda C_d(q)
\]
denote the class-specific Lagrangian, where $G_d(q)$ is
the expected tax gap and $\lambda$ is the multiplier on
the aggregate enforcement budget.

For cutoffs below the frontier, differentiation gives
\[
\frac{d\mathcal L_d(q)}{dq}
=
h_{Y|D}(q\mid d)
\left[
\frac{\lambda(\rho_Y-\rho_D)}{\pi}
-I(q\mid d)
\right].
\]
Since $I(\cdot\mid d)$ is strictly increasing, any interior
stationary point is a local maximum rather than a minimum.
Hence, the preferred cutoff among those below the frontier
lies at a boundary, either $\underline y$ or the limit
as $q$ approaches $\hat y$ from below.

At the global frontier, data verification is no longer
required, so that
\[
C_d(\hat y)
=
\frac{\rho_Y}{\pi}H_{Y|D}(\hat y\mid d).
\]
The tax gap is continuous in the cutoff, while the
enforcement cost falls discontinuously at $\hat y$ by
\[
\frac{\rho_D}{\pi}
\bigl[1-H_{Y|D}(\hat y\mid d)\bigr].
\]
Consequently, the frontier cutoff dominates the limit
from below, leaving only two candidates:
\[
y'(d)\in\{\underline y,\hat y\}.
\]

At the lower cutoff, no profit audits are conducted.
All firms bunch at $\underline y$ and undergo data
verification, giving
\[
C_d(\underline y)=\frac{\rho_D}{\pi}.
\]
The reduction in the tax gap from choosing profit auditing
up to the frontier instead is
\[
G_d(\underline y)-G_d(\hat y)
=
\int_{\underline y}^{\hat y}
I(y\mid d)h_{Y|D}(y\mid d)\,dy.
\]
The corresponding difference in enforcement expenditure is
\[
C_d(\hat y)-C_d(\underline y)
=
\frac{1}{\pi}
\left[
\rho_Y H_{Y|D}(\hat y\mid d)-\rho_D
\right].
\]
Thus,
\begin{align*}
\mathcal L_d(\hat y)-\mathcal L_d(\underline y)
&=
-\int_{\underline y}^{\hat y}
I(y\mid d)h_{Y|D}(y\mid d)\,dy
\\
&\quad+
\frac{\lambda}{\pi}
\left[
\rho_Y H_{Y|D}(\hat y\mid d)-\rho_D
\right].
\end{align*}
Profit auditing up to the global frontier is preferred
to data verification if and only if this difference
is nonpositive, yielding the stated condition.
\end{proof}

\section{Tax-gap comparisons and the value of data}
\label{app:taxgap}

This appendix derives the tax-gap expressions used in
Section~\ref{sec:taxgap} and proves Proposition~\ref{prop:value_data}.
We consider the four enforcement regimes introduced in the main text.

\subsection{Tax gaps across enforcement regimes}

\paragraph{Regime A: no data.}
When the IRS ignores classification data, all firms face the same
profit-audit cutoff $\hat y_A$. The budget constraint is
\begin{equation}
\frac{\rho_Y}{\pi}H_Y(\hat y_A)=B.
\label{eq:budget_A_appendix}
\end{equation}
Equilibrium profit reporting is
\[
x_A(y)=\min\{y,\hat y_A\}.
\]
Firms below the cutoff report truthfully and generate no tax gap, whereas
firms above the cutoff report $\hat y_A$. Hence
\begin{equation}
G^A
=
\int_{\hat y_A}^{\bar y}
\bigl(T(y)-T(\hat y_A)\bigr)\,dH_Y(y).
\label{eq:GA_appendix}
\end{equation}

\paragraph{Regime B: truthful classification data.}
When classification data are truthful, the IRS can condition profit audits
directly on $d$. Let $y'_B(d)$ denote the optimal class-specific cutoff.
Equilibrium reporting is
\[
x_B(y,d)=\min\{y,y'_B(d)\}.
\]
Therefore
\begin{equation}
G^B
=
\int
\int_{y'_B(d)}^{\bar y}
\bigl(T(y)-T(y'_B(d))\bigr)
\,dH_{Y|D}(y\mid d)\,dH_D(d).
\label{eq:GB_appendix}
\end{equation}
The cutoff schedule is chosen subject to
\begin{equation}
\frac{\rho_Y}{\pi}
\int
H_{Y|D}(y'_B(d)\mid d)\,dH_D(d)
=
B.
\label{eq:budget_B_appendix}
\end{equation}

\paragraph{Regime C: strategic classification data with optimal verification.}
Under strategic reporting, let $y'_C(d)$ denote the optimal profit-audit
cutoff and let
\[
\hat y_C:=\sup_d y'_C(d)
\]
denote the associated global verification frontier. Proposition~
\ref{prop:strategic_intensive} and Proposition~
\ref{prop:strategic_extensive} imply the equilibrium reporting rule
\[
x_C(y,d)=\min\{y,y'_C(d)\}.
\]
The resulting tax gap is therefore
\begin{equation}
G^C
=
\int
\int_{y'_C(d)}^{\bar y}
\bigl(T(y)-T(y'_C(d))\bigr)
\,dH_{Y|D}(y\mid d)\,dH_D(d).
\label{eq:GC_appendix}
\end{equation}

The hybrid policy uses profit audits below $y'_C(d)$ and data verification
between $y'_C(d)$ and $\hat y_C$. Its budget constraint is
\begin{equation}
\frac{1}{\pi}
\int
\left[
\rho_D H_{Y|D}(\hat y_C\mid d)
+
(\rho_Y-\rho_D)
H_{Y|D}(y'_C(d)\mid d)
\right]dH_D(d)
=
B.
\label{eq:budget_C_appendix}
\end{equation}
Equivalently, expenditure can be decomposed as
\begin{align}
B
&=
\underbrace{
\frac{\rho_Y}{\pi}
\int
H_{Y|D}(y'_C(d)\mid d)\,dH_D(d)
}_{B_Y:\ \text{profit audits}}
\nonumber\\
&\quad+
\underbrace{
\frac{\rho_D}{\pi}
\int
\left[
H_{Y|D}(\hat y_C\mid d)
-
H_{Y|D}(y'_C(d)\mid d)
\right]dH_D(d)
}_{B_D:\ \text{data verification}}.
\label{eq:budget_C_decomposition_appendix}
\end{align}

Thus $B_D$ is the direct resource expenditure on data verification. Its
opportunity cost in terms of the IRS's objective is instead measured by the
difference between the tax gaps under strategic and truthful data,
\[
C_S:=G^C-G^B.
\]

\paragraph{Regime D: strategic data without verification.}
Finally, suppose that the IRS conditions profit auditing on classification
data but does not verify those data. Firms can then select into the
most leniently enforced class. Let
\[
\tilde y:=\inf_d y'(d)
\]
denote its profit-audit cutoff. The resulting reporting rule is
\[
x_D(y)=\min\{y,\tilde y\},
\]
and the tax gap is
\begin{equation}
G^D
=
\int_{\tilde y}^{\bar y}
\bigl(T(y)-T(\tilde y)\bigr)\,dH_Y(y).
\label{eq:GD_appendix}
\end{equation}

\subsection{Tax-gap rankings}

We first establish the ordering among the optimally designed regimes.

Because Regime A is feasible when data are truthful---the IRS can simply
ignore $d$ and apply the uniform cutoff $\hat y_A$---Regime B optimizes over
a weakly larger set of policies. Hence
\begin{equation}
G^B\le G^A.
\label{eq:GB_GA_appendix}
\end{equation}

Similarly, Regime A is feasible under strategic data: the IRS can ignore
classification reports altogether, in which case no data verification is
required. Therefore
\begin{equation}
G^C\le G^A.
\label{eq:GC_GA_appendix}
\end{equation}

Conversely, the truthful-data problem is a relaxation of the strategic-data
problem. Any policy feasible under strategic reporting must satisfy the
additional incentive constraints required to prevent firms from manipulating
their classification, whereas these constraints are absent under truthful
data. Removing them cannot increase the minimized tax gap. Therefore
\begin{equation}
G^B\le G^C.
\label{eq:GB_GC_appendix}
\end{equation}

Combining the three comparisons gives
\begin{equation}
G^B\le G^C\le G^A.
\label{eq:taxgap_order_optimal}
\end{equation}
Thus
\[
G^A-G^B\ge0,
\qquad
G^C-G^B\ge0,
\qquad
G^A-G^C\ge0.
\]

\subsection{Targeting value and strategic implementation cost}

The gross targeting value of classification data is the improvement that
would be attainable if the classification were truthful:
\begin{equation}
V_T:=G^A-G^B.
\label{eq:VT_appendix}
\end{equation}
Strategic reporting imposes an additional implementation constraint. Its
opportunity cost in terms of the tax gap is
\begin{equation}
C_S:=G^C-G^B\ge0.
\label{eq:CS_appendix}
\end{equation}
It follows immediately that
\begin{align}
G^A-G^C
&=
(G^A-G^B)-(G^C-G^B)
\nonumber\\
&=
V_T-C_S.
\label{eq:value_decomposition_appendix}
\end{align}

Notice that $B_D$ in
\eqref{eq:budget_C_decomposition_appendix} and $C_S$ in
\eqref{eq:CS_appendix} are different objects. The former is the direct
expenditure on data verification, while the latter is its equilibrium
opportunity cost: the increase in the minimum attainable tax gap caused by
having to make data-contingent enforcement incentive compatible.

\subsection{The cost of introducing data}

Suppose that introducing classification data into enforcement entails an
additional fiscal cost $K\ge0$, measured in the same monetary units as the
tax gap. This can capture data administration, information systems, or other
fixed costs that are not included in the enforcement budget $B$.

Relative to ignoring data, adoption generates the net gain
\[
(G^A-G^C)-K.
\]
Hence introducing strategic classification data is optimal if and only if
\[
G^A-G^C\ge K.
\]
Using \eqref{eq:value_decomposition_appendix}, this condition can equivalently
be written as
\begin{equation}
G^A-G^B
\ge
(G^C-G^B)+K.
\label{eq:data_adoption_appendix}
\end{equation}
Thus the gross targeting value must cover both the strategic implementation
cost and the additional cost of introducing the data system.

\subsection{Naive use of manipulable data}

It remains to compare Regime D with the no-data benchmark. Define
\[
F(z)
:=
\int_z^{\bar y}
\bigl(T(y)-T(z)\bigr)\,dH_Y(y).
\]
Then
\[
F'(z)
=
-T'(z)\bigl(1-H_Y(z)\bigr)<0,
\]
so the tax gap is strictly decreasing in the effective profit-audit cutoff.

Under Regime A, the budget is exhausted at $\hat y_A$:
\[
\frac{\rho_Y}{\pi}H_Y(\hat y_A)=B.
\]
Under Regime D, all firms select into the lowest-cutoff class. Feasibility of
the induced policy therefore requires
\[
\frac{\rho_Y}{\pi}H_Y(\tilde y)\le B.
\]
Since $H_Y$ is increasing,
\[
\tilde y\le\hat y_A.
\]
Consequently,
\begin{equation}
G^D
=
F(\tilde y)
\ge
F(\hat y_A)
=
G^A,
\label{eq:GD_GA_appendix}
\end{equation}
with strict inequality whenever $\tilde y<\hat y_A$. Together with
$G^C\le G^A$, this gives
\[
G^D\ge G^A\ge G^C.
\]

\begin{proof}[Proof of Proposition~\ref{prop:value_data}]
Equations \eqref{eq:GB_GA_appendix}--\eqref{eq:GB_GC_appendix} establish
\[
G^B\le G^C\le G^A.
\]
Hence the gross targeting value is $G^A-G^B$, the strategic implementation
cost is $G^C-G^B$, and
\[
G^A-G^C
=
(G^A-G^B)-(G^C-G^B).
\]
If introducing data entails an additional cost $K$, adoption is optimal if
and only if
\[
G^A-G^C\ge K,
\]
which is equivalent to
\[
G^A-G^B\ge(G^C-G^B)+K.
\]
Finally, \eqref{eq:GD_GA_appendix} gives $G^D\ge G^A$ whenever
$\tilde y\le\hat y_A$, while optimality of Regime C gives $G^A\ge G^C$.
This establishes all claims.
\end{proof}

\section{Audit returns, shadow values, and the MVPF}
\label{app:mvpf}

This appendix relates the enforcement index in
Section~\ref{sec:non-strategie} to empirical measures of audit returns and
the marginal value of public funds (MVPF). The key distinction is between
the fiscal return from auditing a marginal taxpayer and the marginal return
from expanding enforcement. In our model, the latter also includes the
information rents eliminated among taxpayers above the enforcement cutoff.

Consider an interior cutoff $y'(d)$. An increase in the class-specific audit
budget expands the set of deterred firms at the cutoff, but also raises the
reported profits of all firms above it. From \eqref{eq:dGdB_gamma},
\[
-\frac{dG(d)}{dB(d)}
=
\frac{\pi}{\rho_Y}
T'(y'(d))
\frac{1-H_{Y|D}(y'(d)\mid d)}
     {h_{Y|D}(y'(d)\mid d)}
=
\frac{\pi}{\rho_Y}I(y'(d)\mid d).
\]
The numerator $1-H_{Y|D}(y'(d)\mid d)$ measures the mass of firms whose
reporting rents are reduced when the cutoff rises, while
$h_{Y|D}(y'(d)\mid d)$ measures the marginal mass that must be brought under
audit coverage. The inverse hazard rate
\[
\gamma(y\mid d)
=
\frac{1-H_{Y|D}(y\mid d)}
     {h_{Y|D}(y\mid d)}
\]
therefore converts the fiscal stakes at the cutoff into the return from
expanding enforcement.

This distinction complements the empirical evidence in
\citet{boning2025welfare}. They estimate both average and marginal fiscal
returns to IRS audits and document substantial heterogeneity across the income
distribution. They also show that the fiscal consequences of an audit extend
beyond the revenue collected in the audited return: individual deterrence
generates additional tax payments for many years after the audit. Our model
captures a different spillover from enforcement. Tightening enforcement at a
cutoff changes the reporting incentives of firms that are not themselves
marginally added to the audit pool. Consequently, observed returns to the
marginal audited taxpayer need not by themselves identify the marginal value
of expanding enforcement.

This distinction has an immediate allocation implication. Let $\kappa$ denote
the shadow value of the aggregate enforcement budget. For classes with
interior cutoffs,
\[
\frac{\pi}{\rho_Y}I(y'(d)\mid d)=\kappa,
\]
and hence, for any two such classes $d$ and $d'$,
\[
T'(y'(d))\gamma(y'(d)\mid d)
=
T'(y'(d'))\gamma(y'(d')\mid d').
\]
Thus heterogeneity in fiscal returns across taxpayer groups need not by itself
indicate misallocation. What must be equalized is the marginal value of
enforcement resources, which also depends on the information rents generated
by the conditional profit distribution.

\subsection{Relation to the MVPF}

The distinction from the MVPF is equally important. The framework of
\citet{hendrensprungkeyser2020} evaluates a marginal policy change by comparing
individuals' willingness to pay with its net fiscal cost. Applying this
framework to tax enforcement, \citet{boning2025welfare} incorporate audit
costs, revenues, deterrence, and taxpayer compliance burdens to evaluate the
welfare consequences of marginal audit expansions.

Our enforcement index answers a different question. Taking the enforcement
budget and the tax schedule as given, it identifies where marginal enforcement
resources have the greatest effect on the tax gap. It is therefore an
allocation statistic rather than a welfare statistic. A complete welfare
analysis could in principle place the enforcement effects characterized here
inside an MVPF calculation, but would additionally require taxpayer welfare,
compliance costs, and the incidence of enforcement.

\subsection{Empirical implementation}

The allocation condition is empirically operational. The tax schedule
identifies $T'(y)$, while administrative data on profits conditional on the
classification variables identify
\[
\gamma(y\mid d)
=
\frac{1-H_{Y|D}(y\mid d)}
     {h_{Y|D}(y\mid d)}.
\]
Together they identify
\[
I(y\mid d)
=
T'(y)\gamma(y\mid d).
\]
At observed interior audit cutoffs, optimal allocation requires the index to
be equalized across data classes. Systematic differences in
$I(y'(d)\mid d)$ therefore provide a diagnostic for the allocation of
profit-audit resources.

This also clarifies what additional information is needed beyond estimates
of audit returns. Evidence such as \citet{boning2025welfare} disciplines the
fiscal consequences of marginal audit expansions; evaluating targeting across
data classes additionally requires the conditional profit distributions that
determine the information rents affected by enforcement.
\section{Comparative statics}
\label{sec:comparative_statics}

This section records several comparative statics implied by the optimal audit
policy. The benchmark results follow directly from the cutoff representation.
Under strategic data reporting, audit costs additionally affect the allocation
between profit audits and data verification and, through this channel, the
global verification frontier.

\subsection{Audit resources and the profit-audit cutoff}

Fix a data class $d$. In the truthful-data benchmark, the within-class budget
constraint is
\[
B(d)
=
\frac{\rho_Y}{\pi}
H_{Y|D}(y'(d)\mid d).
\]
Differentiating with respect to $B(d)$ gives
\begin{equation}
\label{eq:dydB}
\frac{dy'(d)}{dB(d)}
=
\frac{\pi}
{\rho_Y h_{Y|D}(y'(d)\mid d)}
>0.
\end{equation}
A larger class-specific budget therefore extends profit-audit coverage further
into the profit distribution. The response is larger where the conditional
density at the cutoff is low.

Combining \eqref{eq:dydB} with
\[
\frac{dG(d)}{dy'}
=
-
T'(y')
\bigl(1-H_{Y|D}(y'\mid d)\bigr)
\]
yields
\begin{equation}
\label{eq:dGdB}
\frac{dG(d)}{dB(d)}
=
-
\frac{\pi}{\rho_Y}
T'(y'(d))
\gamma(y'(d)\mid d)
=
-
\frac{\pi}{\rho_Y}
I(y'(d)\mid d)
<0.
\end{equation}
Thus, the enforcement index determines the marginal reduction in the tax gap
generated by additional audit resources. Classes with a higher index have a
larger marginal return to profit-audit spending.

\subsection{Profit-audit costs}

Holding the class-specific budget $B(d)$ fixed, differentiation of the budget
constraint with respect to $\rho_Y$ gives
\begin{equation}
\label{eq:dydRhoY_bench}
\frac{dy'(d)}{d\rho_Y}
=
-
\frac{H_{Y|D}(y'(d)\mid d)}
{\rho_Y h_{Y|D}(y'(d)\mid d)}
<0.
\end{equation}
Higher profit-audit costs therefore reduce audit coverage. The magnitude of
the response depends on the mass already covered relative to the density at
the cutoff.

Under strategic data reporting, the marginal cost of increasing a
class-specific profit-audit cutoff depends on whether the class is subject
to data verification. In a verified class, raising the cutoff replaces data
verification with profit auditing, so its incremental resource cost is
$\rho_Y-\rho_D$. In an unverified class, the marginal cost is $\rho_Y$.
For a verified class with an interior cutoff, the optimality condition
\[
I(y'(d)\mid d)
=
\lambda\frac{\rho_Y-\rho_D}{\pi}
\]
shows that, holding the multiplier $\lambda$ and verification status fixed,
a larger $\rho_Y-\rho_D$ lowers the cutoff when $I(y\mid d)$ is decreasing.
This is a conditional response: the optimal multiplier and the set of
verified classes may also change.

\subsection{Data-verification costs and the global frontier}

Strategic reporting introduces an additional cost margin because all
reports above the profit-audit cutoff in a verified class are subject to
data verification. Write
\[
q(d)=\mathbf{1}_{\{y'(d)<\hat y\}}
\]
for the indicator that class $d$ is verified. Total enforcement expenditure is
\begin{equation}
\label{eq:budget_cs_strategic}
B
=
\frac{1}{\pi}
\int
\left[
\rho_Y H_{Y|D}(y'(d)\mid d)
+
\rho_D\bigl(1-H_{Y|D}(y'(d)\mid d)\bigr)q(d)
\right]dH_D(d).
\end{equation}
Thus, unlike a policy that verifies only reports between $y'(d)$ and
$\hat y$, the verification expenditure depends on the entire upper tail
of reports in each verified class.

Holding the cutoff schedule and verification status fixed, the direct
resource effect of a higher verification cost is
\[
\frac{\partial B}{\partial\rho_D}
=
\frac{1}{\pi}
\int
\bigl[1-H_{Y|D}(y'(d)\mid d)\bigr]q(d)\,dH_D(d)
\geq 0,
\]
with strict inequality whenever verified reports have positive mass.
Likewise, the marginal expenditure from raising a cutoff within a class,
without changing its verification status, is
\[
\frac{\partial B}{\partial y'(d)}
=
\frac{h_{Y|D}(y'(d)\mid d)}{\pi}
\bigl(\rho_Y-\rho_D q(d)\bigr),
\]
where the expression is understood as the class-specific integrand of
the aggregate budget (and is weighted by the class probability for a
discrete distribution of classes).

The global frontier $\hat y$ determines which classes require
verification, but it is not an upper limit on the reports verified in
those classes. Consequently, holding the cutoff schedule and the set of
verified classes fixed, expenditure has no direct dependence on
$\hat y$. When the frontier crosses a class-specific cutoff, verification
status changes and the budget can change discontinuously. It would
therefore be incorrect to obtain the response of the optimal frontier to
$\rho_D$ by differentiating the budget constraint with respect to
$\hat y$ while holding the cutoffs fixed.

In the full optimization problem, a change in verification costs may
alter the cutoffs, the verification status of classes, and the optimal
global frontier. The resulting comparative statics are therefore
conditional on the prevailing enforcement regime; in particular, the
optimal policy may switch between differentiated enforcement, partial
pooling, and ignoring classification data.

\begin{remark}[Comparative statics and the enforcement index]
The comparative statics emphasize two distinct roles of the profit
distribution. First, the density at the cutoff determines how far a given
change in audit resources moves the enforcement boundary. Second, the
enforcement index
\[
I(y\mid d)=T'(y)\gamma(y\mid d)
\]
determines the resulting marginal reduction in the tax gap.

Under strategic reporting, profit auditing in a verified class replaces
data verification, giving it incremental resource cost
$\rho_Y-\rho_D$. Verification itself applies to the entire upper tail
above the class-specific cutoff. Its cost therefore affects not only the
allocation between enforcement instruments but also which classes are
verified and how far their profit-audit cutoffs can be differentiated.
\end{remark}

\section{Analytical examples}
\label{app:calibration}

This appendix provides analytical examples of the enforcement index
and the intensive- and extensive-margin enforcement regimes characterized
in the paper. The exercise is not intended as a structural calibration.
Rather, it indicates how empirically meaningful differences in profit
distributions and tax schedules map into optimal enforcement.

We consider three data classes, denoted by
$d_S,d_M,d_L$, corresponding loosely to small, medium, and large firms.
Representative profit levels are normalized to
\[
y_S=1,\qquad y_M=10,\qquad y_L=50.
\]
Throughout, the enforcement index is
\[
I(y\mid d)
=
T'(y)\gamma(y\mid d),
\qquad
\gamma(y\mid d)
=
\frac{1-H_{Y|D}(y\mid d)}
     {h_{Y|D}(y\mid d)}.
\]

The exercise emphasizes that the shape of the conditional profit
distribution determines whether enforcement operates primarily on the
intensive or extensive margin.

\subsection{Intensive-margin illustration: Weibull profits}

Suppose profits conditional on data class $d$ follow a Weibull distribution
with shape parameter $k>1$ and scale parameter $\lambda(d)>0$:
\[
H_{Y|D}(y\mid d)
=
1-\exp\left[
-\left(\frac{y}{\lambda(d)}\right)^k
\right],
\qquad y\geq0.
\]
The inverse hazard rate is
\[
\gamma(y\mid d)
=
\frac{\lambda(d)^k}{k y^{k-1}},
\]
which is strictly decreasing in $y$. Under a linear tax schedule,
$T'(y)=\tau$, the enforcement index is therefore
\begin{equation}
I(y\mid d)
=
\tau
\frac{\lambda(d)^k}{k y^{k-1}},
\label{eq:index_weibull_calibration}
\end{equation}
and is decreasing within every data class. This is the intensive-margin
case characterized in Proposition~\ref{prop:strategic_intensive}.

For example, set
\[
k=2,
\qquad
\tau=0.25,
\]
and choose the scale parameter in each class so that its mean equals the
representative profit $y_d$. Since the mean of a Weibull random variable is
$\lambda(d)\Gamma(1+1/k)$, this gives
\[
\lambda(d)
=
\frac{y_d}{\Gamma(1+1/k)}.
\]
Evaluating the enforcement index at a common relative position in each
class illustrates how differences in the conditional profit distribution
translate into differences in marginal enforcement returns.

More generally, Proposition~\ref{prop:strategic_intensive} implies that
interior profit-audit cutoffs satisfy
\[
I(y'(d)\mid d)=\sigma.
\]
For the Weibull specification with $k=2$, this gives the explicit cutoff
\begin{equation}
y'(d)
=
\frac{\tau\lambda(d)^2}{2\sigma}.
\label{eq:weibull_cutoff_calibration}
\end{equation}
Thus, classes with larger scale parameters receive higher profit-audit
cutoffs at a given shadow value $\sigma$. The numerical exercise therefore
shows the intensive-margin targeting mechanism directly: information
about the conditional profit distribution generates differentiated audit
coverage across data classes.

\subsection{Extensive-margin illustration: Pareto profits}

As a contrasting case, suppose profits conditional on class $d$ follow a
Pareto distribution with shape parameter $\alpha>1$ and scale parameter
$\theta(d)>0$:
\[
H_{Y|D}(y\mid d)
=
1-
\left(\frac{\theta(d)}{y}\right)^\alpha,
\qquad
y\geq\theta(d).
\]
Its inverse hazard rate is
\[
\gamma(y\mid d)
=
\frac{y}{\alpha}.
\]
Under a linear tax schedule,
\[
I(y\mid d)
=
\tau\frac{y}{\alpha},
\]
which is increasing in profits. The Pareto specification therefore provides
a simple illustration of the extensive-margin case in
Proposition~\ref{prop:strategic_extensive}.

Conditional on a global verification frontier $\hat y$, the authority does
not optimally choose an interior profit-audit cutoff in the
extensive-margin case. Instead,
\[
y'(d)\in\{\underline y,\hat y\},
\]
subject to the support of the conditional profit distribution. A class
with $y'(d)=\hat y$ receives profit audits below the frontier and does
not require data verification. A class assigned the lower cutoff is
instead sustained by verifying all reports above that cutoff, including
reports exceeding $\hat y$.

The selection of classes for profit auditing consequently reflects both
their enforcement returns and the full resource cost of verification.
For a candidate class with lower cutoff $a(d)$, replacing verification
above $a(d)$ by profit auditing up to $\hat y$ changes enforcement
expenditure by
\[
\frac{1}{\pi}\left[
\rho_Y\bigl(H_{Y|D}(\hat y\mid d)-H_{Y|D}(a(d)\mid d)\bigr)
-
\rho_D\bigl(1-H_{Y|D}(a(d)\mid d)\bigr)
\right],
\]
provided the alternative with profit auditing up to $\hat y$ does not
require verification. The second term includes the entire upper tail,
not merely reports below the frontier. Thus, although the increasing
enforcement index still favors concentrating profit audits on classes
with high enforcement returns, class selection must account for the
full verification expenditure.

The Pareto example therefore showcases a qualitatively different use of
classification data. Rather than determining how far profit auditing
extends within every class, data help determine which classes receive
profit audits and which are sustained through data verification.

\subsection{Tax progressivity}

The tax schedule provides a second source of heterogeneity in enforcement
returns. Since
\[
I(y\mid d)
=
T'(y)\gamma(y\mid d),
\]
higher marginal tax rates increase the enforcement value of the corresponding
profit levels.

For instance, consider the stylized three-bracket schedule
\[
T'(y)
=
\begin{cases}
0.15, & y<y_M,\\[3pt]
0.25, & y_M\leq y<y_L,\\[3pt]
0.30, & y\geq y_L.
\end{cases}
\]
Relative to a linear tax rate of $25$ percent, this schedule lowers the
enforcement index at low profits and raises it at high profits:
\[
\frac{I_{\mathrm{prog}}(y\mid d)}
     {I_{\mathrm{linear}}(y\mid d)}
=
\frac{T'_{\mathrm{prog}}(y)}{0.25}.
\]
Hence
\[
\frac{I_{\mathrm{prog}}(y_S\mid d)}
     {I_{\mathrm{linear}}(y_S\mid d)}
=
0.6,
\qquad
\frac{I_{\mathrm{prog}}(y_L\mid d)}
     {I_{\mathrm{linear}}(y_L\mid d)}
=
1.2.
\]
Progressivity therefore rotates enforcement incentives toward high-profit
firms. In the intensive-margin case this changes the relative audit cutoffs
across classes; in the extensive-margin case it changes the ranking of
classes according to their average enforcement gains.

\section{Quantitative illustration}
\label{app:quantitative}

This section provides a quantitative illustration of the model. The exercise
corroborates the main theoretical results and assesses their quantitative
importance. In particular, we quantify how much classification data can
improve the targeting of profit audits, how much of this gain is lost when
the data are strategically reported, and how the resulting value of data
depends on their informativeness and the cost of verification.

It should be noted that the exercise is not intended as a structural estimation of a particular tax
administration. We instead discipline the main parameters using estimates
from the empirical tax literature and examine the robustness of the results
over plausible parameter ranges. This allows us to put magnitudes on the
mechanisms characterized in the paper while retaining the transparency of
the theoretical model.

\subsection{Quantitative environment}

We take the intensive-margin environment as our baseline setting. Conditional on data
class $d$, profits follow a Weibull distribution,
\[
H_{Y|D}(y\mid d)
=
1-\exp\left[
-\left(\frac{y}{\lambda(d)}\right)^k
\right],
\qquad y\geq0,
\]
with $k>1$. Under a linear tax schedule $T(y)=\tau y$, the enforcement index
is
\[
I(y\mid d)
=
\tau\frac{\lambda(d)^k}{k y^{k-1}},
\]
which is strictly decreasing in profits. Hence optimal enforcement follows
the intensive-margin characterization of
Proposition~\ref{prop:strategic_intensive}.

We consider three equally likely data classes,
\[
d\in\{0,1,2\},
\qquad
\Pr(D=d)=\frac{1}{3},
\]
and parameterize the scale of the conditional profit distribution as
\[
\lambda(d;\beta)
=
\lambda_0
\frac{\exp[\beta(d-1)]}
{\frac{1}{3}(e^{-\beta}+1+e^\beta)}.
\]
The parameter $\beta\geq0$ controls heterogeneity in the conditional profit
distribution across data classes. The normalization keeps average profits
fixed as $\beta$ varies. When $\beta=0$, profits have the same distribution
in every class and classification data have no targeting value. As $\beta$
increases, the conditional profit distributions become more differentiated,
increasing the potential value of conditioning enforcement on $d$.

We set the linear corporate tax rate to $\tau=0.25$, a representative
statutory corporate tax rate in advanced economies \citep{oecdtax2023}.
We normalize the cost of a profit audit to $\rho_Y=1$ and parameterize the
relative cost of data verification by
\[
r_D:=\frac{\rho_D}{\rho_Y}.
\]
Because the two enforcement technologies are difficult to map directly into
existing administrative cost estimates, we treat $r_D$ as a sensitivity
parameter. Our baseline uses a relatively inexpensive data-verification
technology and we report results over a broad range of values.

Empirical evidence suggests substantial heterogeneity in both enforcement
returns and responses to classification thresholds. \citet{boning2025welfare},
for example, estimate total revenue returns to audit spending that differ by
more than a factor of two across taxpayer groups, while
\citet{almunia2018under} document economically significant reporting responses
around the Spanish Large Taxpayer Unit threshold. We use this evidence to
discipline the range of heterogeneity across data classes rather than to
estimate $\beta$ directly.

Our baseline setting uses $k=2$, as in the analytical Weibull illustration above.
We normalize the aggregate enforcement budget to $B=0.05$. Together with
$\pi=2$ and $\rho_Y=1$, this generates a nondegenerate allocation of audit
coverage across classes. The baseline is chosen to generate meaningful
cross-class heterogeneity while remaining in the interior hybrid region in
which both profit auditing and data verification are used. None of the main
conclusions relies on this particular choice, as shown by the sensitivity
exercises below.

In summary, our baseline sets
\[
k=2,\quad \lambda_0=1,\quad \tau=0.25,\quad \pi=2,
\quad
\rho_Y=1,\quad r_D=0.03,\quad B=0.05,\quad \beta=0.35.
\]
The scale normalizations $\lambda_0=\rho_Y=1$ are without loss for the
comparative exercise. The value $\tau=0.25$ is representative of statutory
corporate tax rates, while $\pi=2$ follows the baseline used in related
numerical tax-enforcement exercises. The baseline values of $r_D$ and $B$
generate an interior hybrid policy with both profit auditing and data
verification.

\begin{table}[H]
\centering
\caption{Baseline parameters}
\label{tab:quantitative_parameters}
\begin{tabular}{lcccc}
\toprule
Parameter & Symbol & Baseline & Range & Discipline \\
\midrule
Weibull shape
    & $k$ & $2$ & $1.5$--$3$
    & analytical benchmark \\
Mean-profit scale
    & $\lambda_0$ & $1$ & --- 
    & normalization \\
Class heterogeneity
    & $\beta$ & $0.35$ & $0$--$0.70$
    & empirical literature \\
Tax rate
    & $\tau$ & $0.25$ & --- 
    & \citet{oecdtax2023} median \\
Penalty parameter
    & $\pi$ & $2$ & $1.5$--$3$
    & numerical benchmark \\
Profit-audit cost
    & $\rho_Y$ & $1$ & --- 
    & normalization \\
Relative verification cost
    & $r_D$ & $0.03$ & $0$--$0.20$
    & sensitivity parameter \\
Enforcement budget
    & $B$ & $0.05$ & $0.03$--$0.08$
    & interior hybrid allocation \\
\bottomrule
\end{tabular}
\end{table}

\subsection{Enforcement regimes and value of data}

For each parameterization, we solve the four enforcement regimes introduced
in Section~\ref{sec:taxgap}. Regime $A$ ignores classification data and
chooses a common profit-audit cutoff. Regime $B$ conditions profit audits on
truthful classification data. Regime $C$ allows classification data to be
strategically reported and optimally combines profit audits with data
verification. Regime $D$ conditions profit auditing on classification data
without verifying those data.

Let the corresponding optimized tax gaps be
\[
G^A,\qquad G^B,\qquad G^C,\qquad G^D.
\]
As established in Section~\ref{sec:taxgap},
\[
G^B\leq G^C\leq G^A\leq G^D.
\]
The quantitative exercise therefore focuses not on this ranking itself, but
on the magnitudes of the wedges between the regimes.

We decompose the value of classification data as
\[
V_T:=G^A-G^B,
\qquad
C_S:=G^C-G^B,
\qquad
V_N:=G^A-G^C=V_T-C_S.
\]
Here $V_T$ is the gross targeting value of classification data, $C_S$ is the
implementation cost generated by strategic reporting, and $V_N$ is the net
enforcement value of strategically reported data.

A useful summary statistic is
\[
R_V
:=
\frac{V_N}{V_T}
=
\frac{G^A-G^C}{G^A-G^B},
\]
whenever $V_T>0$. The statistic $R_V$ measures the share of the potential
targeting value of classification data that survives once strategic
manipulation and the associated verification requirements are taken into
account.

\subsection{Baseline results}

Table~\ref{tab:quantitative_baseline} reports the optimized tax gaps and
enforcement allocations under the four regimes. The truthful-data benchmark
isolates the targeting value created by conditioning profit audits on the
classification variable. Moving from Regime $B$ to Regime $C$ then measures
the amount of this targeting gain that must be sacrificed to make the
classification incentive compatible.

\begin{table}[H]
\centering
\caption{Quantitative illustration: baseline results}
\label{tab:quantitative_baseline}
\begin{tabular}{lcccc}
\toprule
& Regime $A$ & Regime $B$ & Regime $C$ & Regime $D$ \\
\midrule
Tax gap & 0.151 & 0.140 & 0.148 & 0.186 \\
Profit-audit expenditure & 0.050 & 0.050 & 0.045 & 0.013 \\
Data-verification expenditure & 0 & 0 & 0.005 & 0 \\
\midrule
Profit-audit cutoff, $d=0$ & 0.290 & 0.144 & 0.146 & 0.144 \\
Profit-audit cutoff, $d=1$ & 0.290 & 0.290 & 0.386 & 0.144 \\
Profit-audit cutoff, $d=2$ & 0.290 & 0.585 & 0.386 & 0.144 \\
Global frontier & 0.290 & --- & 0.386 & --- \\
\bottomrule
\end{tabular}
\end{table}

The baseline results illustrate all three forces emphasized by the theory.
First, truthful classification data improve the targeting of profit audits.
Relative to ignoring data, the tax gap falls from $G^A=0.151$ to
$G^B=0.140$. The corresponding gross targeting value of data is
\[
V_T=G^A-G^B=0.011,
\]
or 7.3 percent of the no-data tax gap.

Second, strategic data reporting limits but does not eliminate this targeting
value. Under optimal verification, the tax gap is $G^C=0.148$, implying an
implementation cost of approximately $C_S=0.0075$. The remaining net value is
approximately $V_N=0.0037$, or 32.9 percent of the gross targeting value.
Thus strategic reporting dissipates about 67.1 percent of the potential
benefit of classification data.

The enforcement allocation reveals how this loss arises. With truthful
data, the entire enforcement budget is devoted to profit audits, with cutoffs
\[
\bigl(y'_B(0),y'_B(1),y'_B(2)\bigr)=(0.144,0.290,0.585).
\]
With strategic reporting, approximately $B_Y=0.0452$ is devoted to profit
audits and $B_D=0.0048$ to data verification. The optimal cutoffs become
\[
\bigl(y'_C(0),y'_C(1),y'_C(2)\bigr)=(0.146,0.386,0.386).
\]
The two upper classes therefore share a profit-audit cutoff, while the
lowest class is treated differently. Strategic reporting induces partial
pooling rather than a uniform reduction in audit coverage across classes.
Verification absorbs about 9.5 percent of the enforcement budget, while the
loss in targeting value is substantially larger. This illustrates how
verification expenditure and the incentive constraints associated with
strategic reporting jointly restrict differentiated enforcement.

Finally, using manipulable classification data without verification performs
substantially worse than ignoring the data. Under Regime $D$, firms select
into the most leniently enforced class, yielding $G^D=0.186$. The resulting
tax gap is about 23 percent larger than under Regime $A$. Hence, the baseline
describes both sides of the value of data: classification can materially
improve enforcement when it is used together with appropriate verification,
but naive reliance on manipulable data can make enforcement less effective.

\subsection{The informativeness of classification data}

We next vary $\beta$, which determines how strongly the conditional profit
distribution varies across data classes. This exercise isolates the central
tradeoff in the value of classification data.

When $\beta=0$, the conditional profit distributions coincide. The IRS then
has no reason to differentiate profit-audit coverage across data classes, so
classification data have no targeting value:
\[
V_T=V_N=C_S=0.
\]
As $\beta$ increases, classification becomes more useful for targeting
profit audits. At the same time, the resulting differentiation in enforcement
creates stronger incentives to manipulate classification reports and therefore
increases the need for data verification.

\begin{figure}[H]
\centering
\includegraphics[width=0.8\textwidth]
    {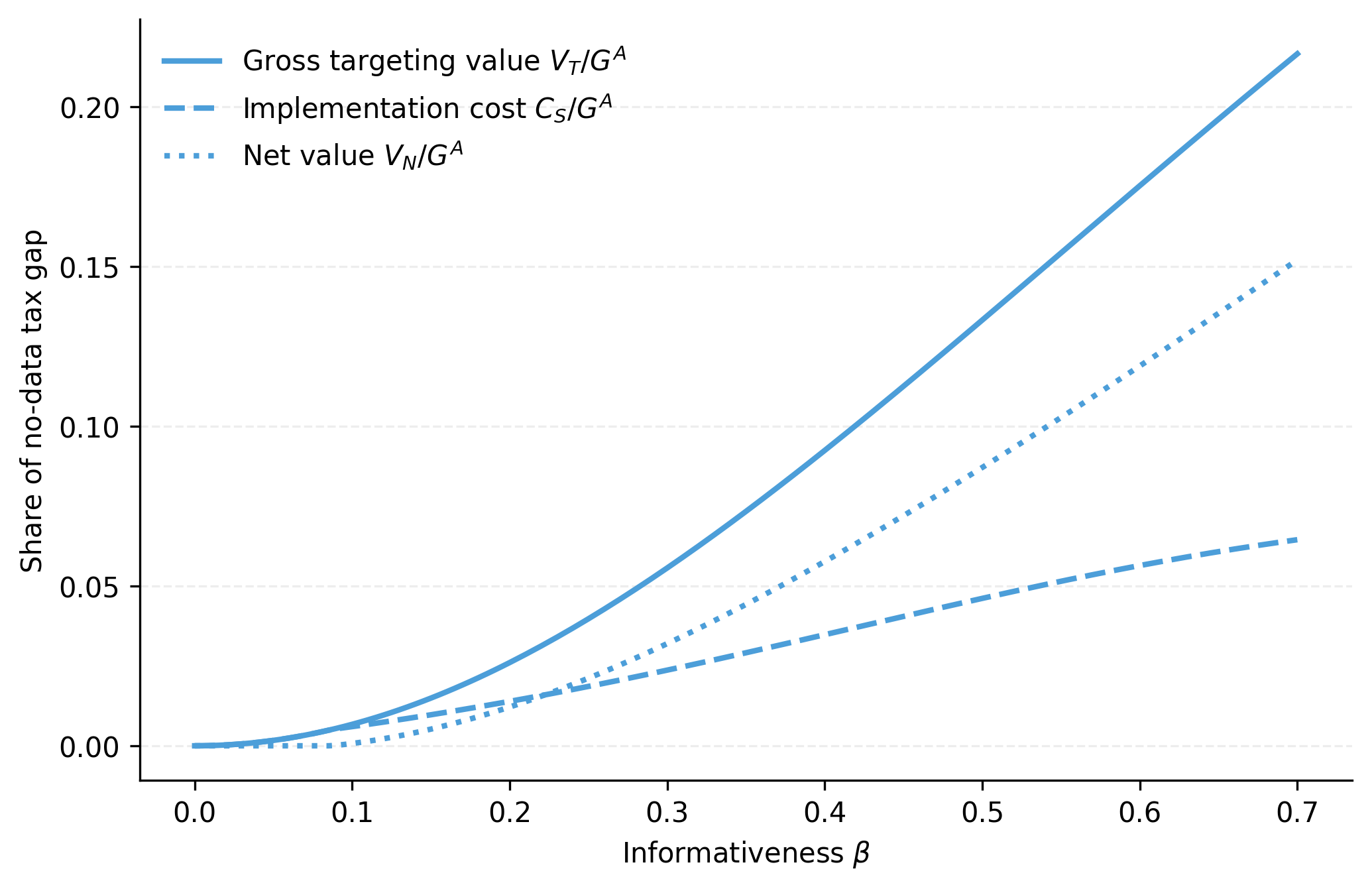}
\caption{The value of data and their informativeness}
\label{fig:value_informativeness}
\end{figure}

Figure~\ref{fig:value_informativeness} plots
\[
\frac{V_T}{G^A},
\qquad
\frac{C_S}{G^A},
\qquad
\frac{V_N}{G^A}
\]
as functions of $\beta$. Normalizing by the no-data tax gap makes the
magnitudes comparable across parameterizations.

Greater heterogeneity increases the potential value of targeting profit
audits, but strategic reporting can eliminate that value at low levels of
informativeness. At $\beta=0.20$, the gross targeting value is approximately
2.6 percent of the no-data tax gap, yet optimal strategic enforcement ignores
data. At the baseline $\beta=0.35$, the gross targeting value is 7.3 percent,
of which 4.9 percentage points are dissipated by strategic implementation,
leaving a net gain of 2.4 percent. At $\beta=0.70$, the corresponding gross,
implementation, and net values are 21.6, 5.2, and 16.5 percent.

Near $\beta=0$, however, the value of differentiated enforcement emerges
more slowly than its verification requirement. To see this, write
$z=d-1\in\{-1,0,1\}$. The normalized scale parameter satisfies
\[
\lambda(d;\beta)
=
\lambda_0
\left[
1+z\beta+O(\beta^2)
\right].
\]
Under truthful data, an interior cutoff satisfies
\[
I(y'_B(d)\mid d)=\kappa.
\]
For Weibull profits this gives
\[
y'_B(d)
=
\left(
\frac{\tau\lambda(d)^k}{k\kappa}
\right)^{1/(k-1)}.
\]
Because the first-order changes in $\lambda(d)$ average to zero across the
three symmetric classes, the budget condition implies $\kappa'(0)=0$.
Therefore,
\[
y'_B(d)
=
y'_0
\left[
1+\frac{k}{k-1}(d-1)\beta+O(\beta^2)
\right].
\]
The dispersion in class-specific cutoffs is consequently first order in
$\beta$. For a differentiated cutoff allocation with a first-order spread,
verification between local cutoffs and the global frontier entails expenditure
of order
\[
B_D=O(\beta).
\]

By contrast, when $\beta=0$ the conditional profit distributions coincide
and the uniform no-data allocation is already optimal. The envelope theorem
therefore implies that a symmetric perturbation generates no first-order
targeting gain:
\[
V_T=G^A-G^B=O(\beta^2).
\]
Hence, for sufficiently small $\beta$, the second-order targeting gain cannot
justify the first-order verification requirement of differentiated
enforcement, and the IRS optimally ignores classification data. In the
baseline parameterization, differentiated strategic-data enforcement becomes
optimal at approximately
\[
\beta^*\approx0.24.
\]

Once this threshold is crossed, the targeting value can exceed the cost of
implementing differentiated enforcement. The retained share
\[
R_V=\frac{V_N}{V_T}
\]
is zero at $\beta=0.20$, approximately 0.33 in the baseline, and about
0.76 at $\beta=0.70$. Thus greater informativeness can eventually make
differentiated enforcement worthwhile despite the verification requirement.

\subsection{The cost of data verification}

We next vary the relative cost of data verification,
\[
r_D=\frac{\rho_D}{\rho_Y}.
\]
The truthful-data benchmark $G^B$ and the no-data benchmark $G^A$ are
independent of $\rho_D$. Hence the gross targeting value
\[
V_T=G^A-G^B
\]
is constant in $r_D$, and changes in the value of strategic data operate
entirely through their implementation cost.

\begin{figure}[H]
\centering
\includegraphics[width=0.8\textwidth]
    {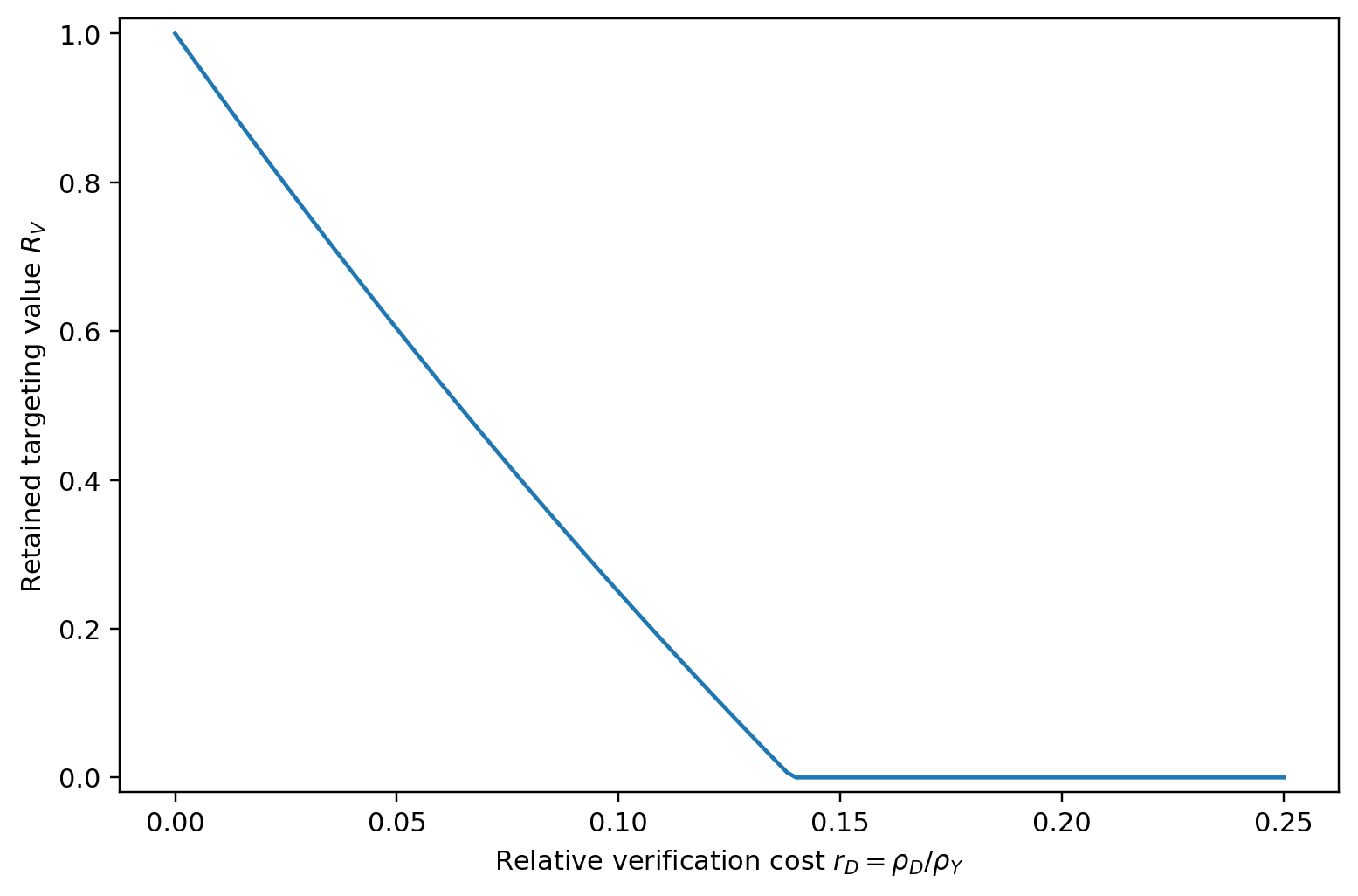}
\caption{The value of data and the cost of verification}
\label{fig:value_verification_cost}
\end{figure}

Figure~\ref{fig:value_verification_cost} plots the retained share
\[
R_V=\frac{G^A-G^C}{G^A-G^B}
\]
as a function of $r_D$.

When data verification is costless, strategic classification can be
implemented without sacrificing profit-audit resources, so
\[
G^C=G^B
\qquad\text{and}\qquad
R_V=1.
\]
As verification becomes more costly, the implementation cost rises and the
net value of classification data falls. At $r_D=0.02$, approximately 53.5
percent of the gross targeting value is retained. At the baseline
$r_D=0.03$, this share is 32.9 percent, while at $r_D=0.05$ it is about
9.2 percent.

The retained share reaches zero as the cost of verification rises beyond
approximately $r_D=0.057$ in the numerical grid. The IRS then optimally
disregards classification data, yielding $G^C=G^A$ and $R_V=0$.
This illustrates the adoption criterion in
Corollary~\ref{coro:value_data}: classification data are worth using only
when the targeting gains exceed the cost of sustaining truthful
classification.

The gross targeting value remains 7.3 percent of the no-data tax gap
throughout this exercise. The implementation cost rises from 3.4 percent
of that gap at $r_D=0.02$ to 4.9 percent at the baseline and 6.7 percent
at $r_D=0.05$. Correspondingly, the net value falls from 3.9 to 2.4
and then to 0.7 percent of the no-data tax gap.

\subsection{Robustness}

Table~\ref{tab:quantitative_robustness} varies the Weibull shape parameter,
the enforcement budget, and the penalty parameter around the baseline. The
main quantitative conclusions are robust across all specifications.

\begin{table}[H]
\centering
\caption{Robustness of the quantitative illustration}
\label{tab:quantitative_robustness}
\begin{tabular}{lcccc}
\toprule
Specification & $V_T/G^A$ & $C_S/G^A$ & $V_N/G^A$ & $R_V$ \\
\midrule
Baseline & 0.073 & 0.049 & 0.024 & 0.329 \\
$k=1.5$ & 0.049 & 0.037 & 0.011 & 0.229 \\
$k=3$ & 0.141 & 0.062 & 0.079 & 0.562 \\
$B=0.03$ & 0.053 & 0.048 & 0.005 & 0.090 \\
$B=0.08$ & 0.101 & 0.043 & 0.058 & 0.572 \\
$\pi=1.5$ & 0.061 & 0.048 & 0.013 & 0.210 \\
$\pi=3$ & 0.096 & 0.044 & 0.052 & 0.540 \\
\bottomrule
\end{tabular}
\end{table}

Across these robustness exercises, the retained share $R_V$ ranges from
approximately 0.09 to 0.57, compared with 0.33 in the baseline. Strategic
reporting therefore absorbs a substantial fraction of the gross targeting
value, although the extent varies considerably across specifications.

Changes in the Weibull shape parameter have a large quantitative effect.
At $k=3$, the gross targeting value is 14.1 percent of the no-data tax gap,
of which approximately 56 percent survives strategic reporting. At $k=1.5$,
the gross targeting value is 4.9 percent and the retained share is about
23 percent.

The value of classification data also rises with the enforcement budget
and with the penalty parameter in these exercises. Increasing the budget
from $B=0.03$ to $B=0.08$ raises the net value from 0.5 to 5.8 percent of
the no-data tax gap, and the retained share from 0.09 to 0.57. Increasing
$\pi$ from 1.5 to 3 raises the net value from 1.3 to 5.2 percent and
the retained share from 0.21 to 0.54. In all reported specifications,
optimally verified strategic-data enforcement remains strictly preferable
to ignoring classification data.

The results indicate that the baseline magnitudes are sensitive to the
parameters governing enforcement returns and resources, while the
qualitative distinction between gross targeting gains and the cost of
strategic implementation persists.

As an additional distributional robustness check, we replace the Weibull
distribution by a lognormal conditional profit distribution, preserving the
same class-specific mean structure and baseline enforcement parameters.
Across log standard deviations between $0.35$ and $1$, the retained share
$R_V$ ranges from approximately $0.04$ to $0.62$. The value of strategic
data therefore remains positive in these specifications, although the
retained share is highly sensitive to the distributional parameters.

The baseline uses three equally likely data classes for transparency. As a
robustness check, we allow the class shares to be asymmetric. Holding the
high-return class at $p_2=0.10$, we vary the relative masses of the remaining
two classes. The value of strategically reported classification data is
non-monotone in the class composition. Over an intermediate range of class
shares, differentiated enforcement remains worthwhile, but when the
classification becomes sufficiently unbalanced, the verification requirement
exhausts the targeting gain and the authority optimally disregards the data.

\begin{table}[H]
\centering
\caption{Robustness to asymmetric class shares}
\label{tab:class_share_robustness}
\begin{tabular}{lccccccc}
\toprule
& $(p_0,p_1,p_2)$ & $G^A$ & $G^B$ & $G^C$ & $G^D$
& $R_V$ & $B_D/B$ \\
\midrule
Equal shares & $(0.33,0.33,0.33)$
& 0.151 & 0.140 & 0.148 & 0.186 & 32.9\% & 9.5\% \\
Weak asymmetry & $(0.40,0.50,0.10)$
& 0.132 & 0.125 & 0.131 & 0.156 & 9.5\% & 11.3\% \\
Moderate asymmetry & $(0.60,0.30,0.10)$
& 0.123 & 0.117 & 0.123 & 0.141 & 0.0\% & 0.0\% \\
Strong asymmetry & $(0.70,0.20,0.10)$
& 0.119 & 0.113 & 0.119 & 0.133 & 0.0\% & 0.0\% \\
\bottomrule
\end{tabular}
\end{table}

For example, with
\[
(p_0,p_1,p_2)=(0.7,0.2,0.1),
\]
truthful classification reduces the tax gap from $G^A=0.119$ to
$G^B=0.113$. Under strategic reporting, however, optimal enforcement reverts
to the no-data policy, so that
\[
G^C=G^A=0.119.
\]
Hence, in this quantitative illustration with a discrete number of classes, the targeting value of data can
be entirely dissipated by the cost of sustaining truthful classification.

\subsection{Extensive-margin illustration}

The baseline quantitative exercise focuses on the decreasing-index case. To
study the extensive-margin result, we retain the same three data classes,
profit distributions, enforcement costs, and budget, but replace the linear
tax schedule by the progressive schedule
\[
T(y)=\tau y^3.
\]
With $k=2$, the enforcement index becomes
\[
I(y\mid d)
=
\frac{3\tau}{2}\lambda(d)^2 y,
\]
which is strictly increasing in profits. Proposition~\ref{prop:strategic_extensive}
therefore applies.

Under the baseline parameter values, the optimal policy profit-audits only
class $d=2$, up to the global frontier
\[
\hat y=0.714,
\]
while classes $d=0$ and $d=1$ are sustained through data verification. At
this frontier, the average enforcement indices are
\[
\bar I(0;\hat y)=0.073,\qquad
\bar I(1;\hat y)=0.155,\qquad
\bar I(2;\hat y)=0.322.
\]
Thus the class with the highest average enforcement return is selected for
profit auditing, as characterized in Proposition~\ref{prop:strategic_extensive}.

The resulting tax gap is $0.387$. By comparison, profit-auditing classes
$d=1,2$ yields a tax gap of $0.401$, while profit-auditing all three classes
yields $0.407$. The numerical exercise therefore illustrates the extensive-
margin logic: when enforcement returns increase in profits, concentrating
profit audits on the highest-return classes can dominate spreading audit
coverage across all classes, with data verification sustaining the resulting
cross-class differentiation. This mechanism is qualitatively consistent with
the substantial cross-group heterogeneity in fiscal audit returns documented by
\citet{boning2025welfare}, although their empirical return measure is not
identical to the enforcement index used here.

\newpage
\bibliography{bibliography}

\end{document}